\documentclass[11pt]{article}  
 \pdfoutput=1 
\usepackage{jheppub}
\usepackage{titlesec}
\usepackage{tikz, amsmath}  
\usetikzlibrary{calc}
\usepackage{bbm}
\usepackage{xcolor} 
\usepackage{dsfont} 
\usepackage{subfigure} 
\usepackage{hyperref} 
\usepackage{microtype}
\usepackage{caption} 
\usepackage{amsmath}
\usepackage{float}
\usepackage{graphics}  
\usepackage{psfrag}
\usepackage{color}
\usepackage{slashed}
\usepackage{subfigure}
\usepackage{graphicx}
\usepackage{array,multirow} 
\usepackage{amsfonts}
\usepackage{amssymb}
\usepackage{mathtools} 
\usepackage{amsmath}
\usepackage{amsfonts}
\usepackage{mathrsfs}
\usepackage{multirow}
\usepackage{feynmp}
\usepackage{array}
\usepackage{textcomp} 
\usepackage{phaistos} 
\usepackage[utf8]{inputenc}
\usepackage{pifont} 
\usepackage{pgf} 

\usepackage{subfigure}
\usepackage{multirow}

\newcommand{\beq}{\begin{eqnarray}}
\newcommand{\eeq}{\end{eqnarray}}

\newcommand{\bmp}{\noindent\begin{minipage}{16cm}}
\newcommand{\emp}{\end{minipage}\vskip 7mm} 

\usepackage{dcolumn}
\usepackage{bm}
\usepackage{slashed}

\usepackage{epsfig}

\usepackage[ margin=5pt, font=normalsize,labelfont=bf,justification=raggedright]{caption}

\usepackage{hyperref}

\newcommand{\bea}{\begin{eqnarray}}
\newcommand{\eea}{\end{eqnarray}}

\newcommand{\ba}{\begin{eqnarray}}
\newcommand{\ea}{\end{eqnarray}}

\title{\Large Partial compositeness under precision scrutiny}
\author[1]{Haiying Cai,}
\emailAdd{hcai@korea.ac.kr}
\affiliation[1]{Department of Physics, Korea University, Seoul 136-713, Korea}

\author[2,3]{Giacomo Cacciapaglia}
\emailAdd{g.cacciapaglia@ipnl.in2p3.fr}
\affiliation[2]{Institut de Physique des Deux Infinis de Lyon (IP2I), UMR5822, CNRS/IN2P3, F-69622 Villeurbanne Cedex, France}
\affiliation[3]{University of Lyon, Universit\'e Claude Bernard Lyon 1,
F-69001 Lyon, France} 

\abstract{We revisit the impact of top partial compositeness on electroweak precision observables in the misaligned vacuum basis. We identify a new source for $S$ in the singlet mixing case, and for $S$-$T$ in the bi-doublet mixing, stemming from misalignment in the gauge couplings of the top partners. Hence, a positive shift in $T$ can be obtained in both cases, as preferred by the recent CDF measurement of the $W$ mass. These results, obtained for the minimal fundamental coset SU(4)/Sp(4), apply to any composite Higgs model with top partial compositeness.}

\begin{document}
\maketitle

\section{Introduction}

The Brout-Englert-Higgs sector of the Standard Model (SM) \cite{Englert:1964et,Higgs:1964pj} has been introduced to break the electroweak (EW) symmetry and give mass to the $W$ and $Z$ boson, and all elementary fermions. It consists of an EW doublet scalar field that develops a non-zero expectation value on the vacuum, leading to the longitudinal degrees of freedom of the massive gauge bosons and a physical scalar field, the Higgs boson \cite{Higgs:1964pj}.  Since the establishment of the SM as the standard theory for particle interactions, it has been tantalising to replace this sector with a composite one, i.e. a strongly interacting sector where the symmetry breaking is achieved dynamically \cite{Weinberg:1975gm}. The main inspiration for this idea is Quantum Chromodynamics (QCD), another fundamental sector of the SM where dynamical condensation of quarks breaks the chiral symmetry of the model. In QCD, scalar particles emerge as resonances made of quarks (or gluons). A notable example are pions, quark-antiquark bound states that remain lighter than other states as they are pseudo-Nambu-Goldstone bosons of the chiral symmetry breaking.

In composite Higgs models (CHMs), the dynamical symmetry breaking pattern is arranged in such a way that an EW doublet can be constructed out of four pNGBs \cite{Kaplan:1983fs}, hence explaining naturally a mass hierarchy between the Higgs boson and other states appearing in the theory.  This mechanism helps addressing the `Little Hierarchy' problem, i.e. the absence of new physics at the EW scale \cite{Barbieri:2000gf}, while the hierarchy with the GUT or Planck scales is explained dynamically by the absence of fundamental scalars in the theory. Besides the pNGB Higgs, recent models employ partial compositeness \cite{Kaplan:1991dc} to give mass to the SM fermions, most notably to the top quark. In CHMs, and more generally in all dynamical models, fermion masses arise from higher dimensional operators that couple the SM fermion fields to operators from the strong sector. Hence, the emerging effective Yukawa couplings are typically suppressed and much smaller than unity. This may work for all SM fermions except for the top quark, whose mass is of the same order as the EW scale. In partial compositeness, direct Yukawa-like couplings are replaced by linear mixing of the SM top fields with spin-1/2 composite operators. If these operators have suitably large anomalous dimensions, a large Yukawa coupling for the top can be explained.  Note that a conformal behaviour above the condensation scale is required in order to solve the flavour puzzle in these models \cite{Holdom:1981rm}, like in walking Technicolor \cite{Bando:1987we,Foadi:2007ue}.

This class of models received a lot of attention recently, raising as a valid alternative to supersymmetry, hence we will not attempt to summarise the main developments as the are nicely contained in comprehensive reviews \cite{Contino:2010rs,Bellazzini:2014yua,Panico:2015jxa,Cacciapaglia:2020kgq}. 
For the economy of this paper, it suffices to recall that the minimal working symmetry breaking pattern (coset) is SO(5)/SO(4) \cite{Agashe:2004rs}, which provides only an EW doublet in the pNGB sector while preserving custodial symmetry \cite{Georgi:1984af,Agashe:2006at}. This was originally proposed as a model in extra dimensions, following the philosophy of gauge-Higgs unification models \cite{Hosotani:1983xw,Hosotani:2005nz}, connected to 4-dimensional theories via holography \cite{Contino:2003ve}.
Following instead the QCD template, the minimal coset is SU(4)/Sp(4) \cite{Cacciapaglia:2014uja}, arising from fundamental fermions charged under a pseudo-real representation of the confining strong gauge interaction. While this coset contains an extra pNGB besides the Higgs doublet, its minimality can be sought in the field content of the underlying gauge-fermion theory. The minimal model, therefore, consists of a gauged SU(2)$_{\rm FC}$ (where FC stands for fundamental (techni)colour \cite{Cacciapaglia:2020kgq}) with two Dirac fermions in the fundamental representation \cite{Ryttov:2008xe,Galloway:2010bp}. Such theories have the advantage of being studied on the Lattice (see Ref. \cite{Cacciapaglia:2020kgq} for a review). Partial compositeness for the top can be obtained by extending the minimal fermion content of the model \cite{Ferretti:2013kya}, with  the presence of two different irreducible representations allowing to sequester QCD gauge interactions from the EW breaking coset \cite{Barnard:2013zea,Ferretti:2013kya}.

In this work we reconsider the partial compositeness sector in the low energy limit, where the composite operators that couple to the top give origin to spin-1/2 baryon-like resonances, which mix with the elementary SM top fields to give rise to massive states. We focus on the minimal SU(4)/Sp(4) case, while highlighting universal properties that apply in general to any fundamental CHM. 
In particular, we will examine the contribution of the top partners to electroweak precision observables, motivated by the recent release of the $W$ mass measurement from the CDF collaboration at Fermilab's Tevatron \cite{doi:10.1126/science.abk1781}. 
This provides the most precise experimental determination of the $W$ mass:
\begin{equation} \label{eq:CDF}
    \left. M_W \right|_{\rm CDF} = 80,433.5 \pm 6.4_{\rm stat} \pm 6.9_{\rm syst}= 80,433.5 \pm 9.4\, {\rm MeV}  \ ,
\end{equation}
which is in strong tension (of about 7 standard deviations) with the determination from SM fits \cite{ParticleDataGroup:2020ssz}. Even taking into account the impact of higher order QCD corrections \cite{Isaacson:2022rts} and the average with previous measurements, a tension remains. If confirmed, this discrepancy could be a strong hint for the presence of new physics beyond the SM in the EW sector, and reveal the origin of the EW symmetry breaking mechanism. Notably, it corresponds to a positive value of the oblique $T$ parameter \cite{Lu:2022bgw,Strumia:2022qkt,deBlas:2022hdk}. The jury is not out to deliberate yet, nevertheless we will investigate how fundamental CHMs may explain this tension and how this impacts the properties of top partners. The latter are effectively vector-like quarks (VLQs) that mix dominantly to the third generation, and their generic impact on the $W$ mass have been investigated \cite{Cao:2022mif,He:2022zjz}.  The main difference between generic VLQ models and CHMs is twofold: on the one hand, CHMs contain non-linear couplings of the Higgs versus linear Yukawa-like couplings  in VLQ models; on the other hand, the two models are usually studied in different equivalent bases (CHMs naturally have gauge-preserving mixing and absent SM-like Yukawas, while in VLQ models only Higgs induced mixings are included). In our work we consider in detail the effects of the misalignment of the CHM vacuum, which induces divergences in the oblique parameters $S$ and $T$ as well as in the longitudinal $W$ scattering amplitudes. We also include the effect of derivative couplings, which are only present in CHMs where the Higgs arises as a pNGB.

The article is organised as follows: in Sec.~\ref{sec:PC} we summarise the general properties of the effective Lagrangian  to describe CHMs with top partial compositeness. We will particularly stress universal properties, which are model-independent. In the following Sec.~\ref{sec:su4sp4}, we specialise to the minimal coset stemming from fundamental composite models with an underlying gauge-fermion description, based on SU(4)/Sp(4). We discuss the impact of top partial compositeness on precision observables in Sec.~\ref{sec:ewt}, before offering our conclusions in Sec.~\ref{sec:concl}. In the Appendix, we  give  the detail calculation of  $S, T$ and $U$ parameters from vector-like fermions in generic models. In particular,  the old $\psi_+$ function for VLQ models in the literature is generalized.

 \section{Effective models for (top) partial compositeness in CHMs} \label{sec:PC}
The low-energy physics of CHMs can be characterised in terms of the CCWZ construction, allowing to describe the effective interactions of light resonances in an expansion in powers of $\partial/\Lambda$ and $m_{\rm res}/\Lambda$, where $\Lambda$ is the condensation scale of the underlying model. All resonances with $m_{\rm res} \ll \Lambda$ can be included  \cite{Marzocca:2012zn}, and here we will focus on the pNGBs (including the Higgs boson) and top partners. Vector and axial-vector can also be included following the hidden symmetry prescription \cite{Bando:1987br}.

To study the EW symmetry breaking, it is crucial to choose the appropriate vacuum of the theory, which breaks the global symmetry $\mathcal{G}$ to a subgroup $\mathcal{H}$. In this work we chose a vacuum that contains the EW symmetry breaking parameter. The procedure goes as follows: 
\begin{itemize}
\item[1)] We define a vacuum $\Sigma_0$, transforming under an appropriate representation of $\mathcal{G}$, which preserves the EW gauge symmetry. In this way, the broken generators $X^a \subset \mathcal{G}/\mathcal{H}$ and the unbroken ones $S^m \subset \mathcal{H}$ can be assigned well-defined transformation properties under the EW gauge symmetry. Within the broken generators, one can identify 4 that transform as a bi-doublet of the custodial symmetry \cite{Contino:2010rs}, hence sharing the same properties as the Higgs scalar field in the SM.
\item[2)] The generator associated to the Higgs boson, $X^4$, is used to define a $\mathcal{G}$ rotation which misaligns the vacuum along the EW symmetry breaking direction \cite{Kaplan:1983fs,Cacciapaglia:2014uja}:
\begin{equation}
U_\alpha = \exp\left[ i \sqrt{2} \alpha  X^4   \right]\,,
\end{equation}
where $f$ is the decay constant of the pNGBs. Here, $\alpha$ is an angle that encodes the equivalent to the Higgs vacuum expectation value in the SM.
\item[3)] All composite objects used to construct the CCWZ effective Lagrangian are rotated via $U_\alpha$, hence they cannot be decomposed into object transforming under the EW symmetry. The basic CCWZ couplings, however, do not depend on $\alpha$, as the strong sector in isolation is invariant under any $\mathcal{G}$-rotation.
\item[4)] The SM couplings (like EW gauge and top effective Yukawas) are introduced via spurions that explicitly break $\mathcal{G}$. For instance, the EW gauge generators are a subset of $S^m$. Due to the misalignment, the dependence on $\alpha$ only appears via spurion insertions.
\end{itemize}
The main benefit of this vacuum choice is that the dependence on $\alpha$ only appears via spurions, hence all the symmetries of the strong sector are explicitly preserved in the limit where the spurions vanish. This allows to clearly isolate the effects due to the EW misalignment. 
In contrast, a popular vacuum choice is to work in terms of the EW preserving one and then assigning a vacuum expectation value to the pNGB associated to the $X^4$ direction \cite{Contino:2010rs}. In this other approach, one can also define a misalignment angle, which however breaks explicitly some symmetries of the pNGBs, such as the shift symmetry along the $X^4$ direction, even without explicit spurion insertion. Nevertheless, as the two choices are related via a field redefinition, the physics they encode is the same in terms of physical observables.

Once our choice clarified, we can introduce the main building blocks for the effective Lagrangian. The pNGBs $\pi^a$ are encoded in the pion matrix:
\beq
U_\Pi = U_{\alpha}\ \exp\left[ \frac{i \sqrt 2}{f} \sum_{a=1}^{dim(\mathcal{G}/\mathcal{H})} ( \pi^a  X^a )  \right]\ U_{\alpha}^{-1}\,,
\eeq
rotated via the misalignment $U_\alpha$. It transforms non-linearly under the global symmetries as follows: 
\beq
U_\Pi \to g_i U_\Pi h(g_i,\pi_i)^{-1} \,,
\eeq
where $h(g_i,\pi_i)$ is in the rotated unbroken group.
Defining the CCWZ object $i \,U_\Pi^{-1} D_\mu U_\Pi $, one can construct two symbols by projecting it along the broken and unbroken directions, as follows:
\beq
d_\mu &=&   \sum_a\ 2 X^a_\alpha tr \left[ X^a_\alpha U_\Pi^{-1} \left(i \partial_\mu + g_2 W^i_\mu  T_L^i + g_1 B_\mu T_R^3 \right) U_\Pi \right]\,, \\
E_\mu &=& \sum_m\ 2 S^m_\alpha tr \left[ S^m_\alpha U_\Pi^{-1} \left(i \partial_\mu + g_2 W^i_\mu T_L^i + g_1 B_\mu T_R^3 \right) U_\Pi \right] 
\eeq
where $X^a_\alpha = U_\alpha X^a U_\alpha^{-1}$ and $S^m_\alpha = U_\alpha S^m U_\alpha^{-1}$ are the broken and unbroken generators in the rotated vacuum, while the gauge generators $T_{L,R}$ are aligned to the original basis. Equivalently,  the net effect is that the gauge group is backward rotated as $ T_{L/R} \to U^{-1}_\alpha T_{L/R} U_\alpha$.
The CCWZ objects built from the one-form transform as:
\beq
&& E_{\mu}\to h(g_i,\pi_i)\, (E_{\mu}+i\partial_\mu)\, h^{-1}(g_i,\pi_i)\,,
\nonumber \\
&& d_{\mu}\to h(g_i,\pi_i)\, d_{\mu}\, h^{-1}(g_i,\pi_i)\,.
\eeq
We noted that, once expanding them for small misalignment angle $\alpha$, the first two terms are universal as they only depend on  the $SU(2)_L\times SU(2)_R$ symmetry in  $\mathcal{G}/\mathcal{H}$:
\beq
E_\mu  &=& \sum_i^3 \left( g_2 W_\mu^i  T_L^i + g_1 B_\mu T_R^i \delta^{i3} \right) - \sin^2 {\frac{\alpha}{2}}  \sum_i^3 \left( g_2 W_\mu^i -g_1 B_\mu \delta^{i 3 }\right) \left(T_L^i -T_R^i \right) + \cdots  \, \label{Emu}\\
d_\mu 
&=& -  \frac{\sqrt{2 }}{f} \partial_\mu   \Big( h X^4  + \sum_{a=5}^{dim(\mathcal{G}/\mathcal{H})}\eta^a X^a \Big) +  \frac{\sin \alpha}{\sqrt{2}} \sum_{i=1}^3 (g_2 W^i_\mu - g_1 B_\mu \delta^{i 3})  X^i + \cdots      \, \label{dmu}
\eeq
Here, $h$ is the pNGB Higgs while $\eta^a$ represent all the remaining pNGBs.
It can be checked explicitly that the expressions above match the ones obtained in the $SO(5)/SO(4)$ model~\cite{Cai:2013kpa}. 
This structure is universal for a generic CHM where the Higgs  is a bi-doublet $X^a, a = 1,\cdots 4$ under $SU(2)_L\times SU(2)_R$, as a consequence of  group algebra.  In fact the property of custodial symmetry  defines the commutator relation in a  non-minimal $\mathcal{G}/\mathcal{H}$ CHM: 
\beq
&& \left[ T_L^i, T_L^j \right] = i \, \epsilon^{ijk} T_L^k \,,  \quad  \quad \left[ T_R^i, T_R^j \right] = i \,\epsilon^{ijk} T_R^k \,, \quad (i = 1, 2, 3)   \nonumber \\
&&  \left[X^4, T_{L/R} ^i \right]= \pm \frac{i}{2} X^i \,,  \quad \left[ X^4, X^i \right] = - \frac{i}{2} \left(T_L^i -T_R^i \right) \, \label{commutator}
\eeq
that will fix the generators  for  NGBs  (missing in the unitary gauge) eaten by $W, Z$ gauge bosons in a specific model building.
Note that Eq.(\ref{commutator}) forms a closed subgroup. The expansion of  $i U_\Pi^{-1}  D_\mu  U_\Pi$ till the linear order  can  be computed  as:
\beq
-  \frac{\sqrt{2 }}{f} \partial_\mu  \Big( h X^4  + \sum_{a=5}^{dim(\mathcal{G}/\mathcal{H})}\eta^a X^a \Big) + U_\alpha^{-1} \Big( \sum_i^3 g_2 W_\mu^i T_L^i + g_1 B_\mu T_R^3 \Big) U_\alpha \,,  \label{CCWZ}
\eeq
where the first term is in the broken direction and the second term can  split  into the $E_\mu$ and $d_\mu$ parts.  Using the commutators in Eq.(\ref{commutator}),  one can derive that\footnote{We use the Baker–Campbell–Hausdorff formula: $\exp(A) B \exp(-A) = B  + [A,B] + [A,[A,B]]/2!+ \cdots$}:
\beq 
U_\alpha^{-1} T_L^i U_\alpha  &=& T_L^i + \frac{\sin\alpha }{\sqrt{2}}X^i - \sin^2\frac{\alpha}{2} \left( T_L^i -T_R^i \right)  \,,   \nonumber \\ 
U_\alpha^{-1} T_R^i U_\alpha  &=& T_R^i - \frac{\sin\alpha }{\sqrt{2}}X^i + \sin^2\frac{\alpha}{2} \left( T_L^i -T_R^i \right) \,. \label{TLR}
\eeq
Applying Eq.(\ref{TLR}) to Eq.(\ref{CCWZ}),  we immediately obtain the final form of  Eq.(\ref{Emu}-\ref{dmu}).   A direct prediction from this universal structure  is its imprint on the Higgs gauge couplings that can be alternatively  inferred from the infrared construction of pion scattering amplitude \cite{Liu:2018qtb}. However  the higher order expansion of CCWZ object depends on the commutators involving with  the additional pNGB in a non-minimal coset $\mathcal{G}/\mathcal{H}$.

The top partners arise as baryons of the confining theory, hence they are represented by spin-1/2 resonances transforming as representations of the unbroken group $\mathcal{H}$. Typically, they also carry additional SM charges to match those of the elementary top fields: here we will consider additional QCD charges and hypercharge, motivated by the models in Refs~\cite{Ferretti:2013kya,Belyaev:2016ftv}.
As for the pNGBs, in our approach they can be expresses in terms of the EW preserving vacuum, so that the components' EW charges can be clearly identified. Then, they are rotated by $U_\alpha$ to be aligned to the  rotated vacuum.
The SM spinors, instead, are embedded in spurions transforming as incomplete representations of $\mathcal{G}$ and they  can couple to matching baryons.

To be concrete, let's consider the SM  top and bottom are embedded into a two-index representation of $\mathcal{G}$ in a real -- SU(N)/SO(N) -- or pseudo-real -- SU(2N)/Sp(2N) -- coset. The Lagrangian of elementary fields is ${\mathcal L}_{elem} =  \bar {\psi}_{SM}  (i  \partial_\mu + g_2 W_\mu^i \sigma^i/2 + g_1 \hat Y  B_\mu +  g_s G^a_\mu \lambda_a) \gamma^\mu \psi_{SM} $, with $\psi_{SM} \supset \{ \left( t_L, b_L \right)^T, t_R \,, b_R \}$. The two relevant baryons consist in an antisymmetric of $\mathcal{H}$, $\psi_A =  U_{\alpha}\hat \psi_A U_{\alpha}^T$, and an invariant singlet, $\psi_1 =  U_{\alpha}\hat \psi_1 U_{\alpha}^T $, where the hatted baryons are aligned to the EW preserving vacuum, i.e. to the EW symmetry.  We recall that, once misaligned, the components of the baryons do not have well-defined transformation properties under the EW symmetry.
The baryon Lagrangian can be written as:
\beq
\mathcal{L}_{\rm composite} &=& tr \left[ \bar{\psi}_A\ i \not{D} \  \psi_A \right] - M_A\  tr \left[ \bar{\psi}_A  \psi_A \right] + tr [ \bar{\psi}_1\ i \not{D}\ \psi_1] - M_1\ tr [\bar{\psi}_1  \psi_1] + \nonumber \\
&& +  \kappa' \,  tr \left[ \bar{\psi}_A \not{d}\  \psi_A \right]  + \kappa \, \left( tr \left[ \bar{\psi}_A \not{d}\  \psi_1 \right]  + \mbox{h.c.} \right)  \,. \label{eq:Lcomposite}
\eeq
The first 4 terms are the kinetic terms, including the covariant derivatives, while the last two contain the derivative couplings to the pNGB fields. For SU(N)/SO(N),   only the $\kappa'$ term survives  because ${\not d} \psi_1$ is symmetric in SU(N)/SO(N),  leading to  $ \kappa \, tr \left[\bar{\psi}_A \not d \psi_1 \right] =0$. The covariant derivative for the composite baryons, including the additional QCD coupling and hypercharge $X$, reads
\beq
D_\mu  = ( \partial_\mu  - i E_\mu - i g_1 X  B_\mu  - ig_s G^a_\mu \lambda_a  )\,.
\eeq
Note again that, if we turned off the EW gauge couplings, the above Lagrangian would not depend on the misalignment angle $\alpha$. The $\alpha$ dependence also emerges from the couplings of the top fields to the baryons: to construct them, one needs to dress the baryon fields $\psi$ with the pNGB matrix in order to obtain operators that transform linearly under $\mathcal{G}$. Then, $\mathcal{G}$--invariant couplings can be built with the spurions containing the SM top fields. We will discuss an explicit case in the next section.

 \section{The SU(4)/Sp(4) Model} \label{sec:su4sp4}
We will use the SU(4)/Sp(4) CHM with a gauge-fermion underlying description as a template to illustrate the common features emerging from top partial compositeness.  The underlying dynamics  for this minimal CHM  is  a strongly interacting gauge theory with fundamental fermions charged under  Sp(2N)$_{\rm FC} \times$SU(3)$_c \times$SU(2)$_L \times$U(1)$_Y$. The FC-charged Weyl fermions include four QCD-colour singlets $Q$  and two QCD-colour (anti)-triplets  $\chi$, leading to a global symmetry $\mathcal{G}=$SU(4)$\times$SU(6)$\times$U(1). This model was first studied in Ref.~\cite{Barnard:2013zea}, while considerations related to the conformal window sets a preference for $\text{N}=2$ \cite{Ferretti:2016upr}.~\footnote{These Sp(4)$_{\rm FC}$ model has been investigated on the Lattice \cite{Bennett:2017kga,Lee:2018ztv,Bennett:2019jzz,Lee:2019pwp,Bennett:2019cxd,Bennett:2021mbw,Bennett:2022yfa} in order to study its spectrum. Investigations based on Nambu-Jona-Lasinio model \cite{Bizot:2016zyu} and on holography \cite{Erdmenger:2020lvq,Erdmenger:2020flu} have also been performed.} Note also that the same global symmetry could arise from a different dynamics based on SO(11)$_{\rm FC}$ and fermions in the spinorial and fundamental representations: the two underlying dynamics will only be distinguished by the properties of the pseudo-scalar pNGB from the spontaneous breaking of the U(1) global symmetry \cite{Belyaev:2016ftv, Cacciapaglia:2019bqz,BuarqueFranzosi:2021kky}. The model generates two condensates, $\langle QQ \rangle$ and $\langle \chi \chi \rangle$, which break the global symmetry to $\mathcal{H}=$Sp(4)$\times$SO(6). In particular, the $\chi$-condensate generates QCD-coloured pNGBs including an octet and a sextet \cite{Cacciapaglia:2015eqa}. Here we will focus on the EW-coset, driven by the $Q$-condensate.

Below the confinement scale $\Lambda_{\rm FC}$, the antisymmetric condensation of $\langle QQ \rangle$ delivers one Higgs bi-doublet plus one singlet $\eta$ as pNGBs, and a potentially light scalar $\sigma$ in the EW sector. At the low energy, the effective theory of mesons realised as pNGBs in the coset SU(4)/Sp(4) is  described by a nonlinear sigma model \cite{Cacciapaglia:2014uja, Arbey:2015exa}. The broken and unbroken generators for this model are adjusted to suit Eq.(\ref{commutator}) and listed in  Appendix \ref{Appendix2}.  By identifying the SU(2)$_L$ generators as $T_L^i  = S_i$, $i = 1 \cdots 3$ and  $U(1)_R$ as $T_R^3= S_6$, the charge operator is defined $Q = S_3 + S_6 + X$, where the additional hypercharge $X = 2/3$ is carried by the fermions $\chi$. It is assigned to the top partners via the presence of $\chi$ in the baryon operators. The EW-preserving vacuum, which leaves invariant Sp(4)$\supset$SU(2)$_L \times $SU(2)$_R$ in the unrotated basis, is :
\beq
\Sigma_0 = \left( \begin{array}{cc}
i \sigma_2 & 0 \\
0 & - i \sigma_2
\end{array} \right) \,,
\eeq 
while the pion matrix reads :
\beq
\hat \Pi = \sum_{a=1}^5 \pi^a X^a = \left(
\begin{array}{cccc}
 \frac{\eta }{2 \sqrt{2}} & 0 & \frac{G_+}{2} & \frac{G_0-i h}{2 \sqrt{2}} \\
 0 & \frac{\eta }{2 \sqrt{2}} & \frac{i h+G_0}{2 \sqrt{2}} & -\frac{G_-}{2} \\
 \frac{G_-}{2} & \frac{G_0-i h}{2 \sqrt{2}} & -\frac{\eta }{2 \sqrt{2}} & 0 \\
 \frac{i h+G_0}{2 \sqrt{2}} & -\frac{G_+}{2} & 0 & -\frac{\eta }{2 \sqrt{2}} \\
\end{array}
\right)\,,
\eeq
where $G_x$ are the Goldstones eaten by the $W$ and $Z$ bosons in the unitary gauge.
The misalignment  after the EW symmetry breaking is  generated  by a SU(4) rotation matrix :
\beq
U_{\alpha} = \exp{\left[ i \sqrt{2} X^4 \alpha \right] } 
= \left(
\begin{array}{cccc}
 \cos \left(\frac{\alpha }{2}\right) & 0 & 0 & \sin \left(\frac{\alpha }{2}\right) \\
 0 & \cos \left(\frac{\alpha }{2}\right) & -\sin \left(\frac{\alpha }{2}\right) & 0 \\
 0 & \sin \left(\frac{\alpha }{2}\right) & \cos \left(\frac{\alpha }{2}\right) & 0 \\
 -\sin \left(\frac{\alpha }{2}\right) & 0 & 0 & \cos \left(\frac{\alpha }{2}\right) \\
\end{array}
\right)
\eeq
such that $\Sigma_0$ and $\hat \Pi$ can be rotated according to the following rules:
\beq
  \Sigma_\alpha =  U_{\alpha} \Sigma_0 U_{\alpha}^T\,, \quad  \Pi = U_{\alpha} \hat \Pi U_{\alpha}^{-1}  \,.
\eeq
With the building blocks above, we write down the nonlinear sigma field:
\beq
\Sigma =   U_{\Pi} \Sigma_\alpha U_{ \Pi}^T \,, \quad 
\mbox{with}  \quad 
 U_{\Pi}  = \exp{\left[ i  \sqrt{2}\frac{ \Pi}{f}\right]}\,. 
\eeq
Since $U_{\Pi}$ transforms nonlinearly as $ g U_{\Pi}  h^{-1}(\Pi, g)$ and  $\Sigma_\alpha \to h \Sigma_\alpha h^T$,  thus $\Sigma \to g \Sigma g^T$ for $g \in SU(4)$ and  $h \in$ the misaligned Sp(4) group.  The kinetic Lagrangian of pNGB is then constructed as:
\beq
\mathcal{L}_{kin} = \frac{f^2}{8} Tr \left[ (D_\mu \Sigma)^\dag D^\mu \Sigma \right]  
\eeq
with $D_\mu = \partial_\mu - i g_2 W_\mu^i T_L^i - i g_1 B_\mu T_R^3$. The effective Higgs vacuum expectation value is defined as $v = f \sin \alpha$ via the expression for the $W$ and $Z$ masses.

\subsection{Top partners}
Below the confinement scale,  we also consider the spin-$\frac{1}{2}$ resonances made of chimera baryons with one $\chi$ and two $Q$ fermions. Lattice results seem to indicate that such states are rather heavy~\cite{Ayyar:2017qdf,Bennett:2022yfa}, nevertheless we will assume here that there exist a mechanism pushing their mass well below the cut-off of the effective theory.

In the model under consideration, the relevant top partners transform as $(\mathbf{5},  \mathbf{6}) \oplus (\mathbf{1},  \mathbf{6}) $ of the unbroken Sp(4)$\times$SO(6), denoted as $ \hat \psi_5 \oplus \hat \psi_1$. The $\bf 6$ of SO(6) simply implies the presence of a triplet and an anti-triplet of QCD, making each baryon a massive vector-like quark. In the EW sector, and around the unrotated vacuum, the baryons' components read:
\beq
\hat \psi_5 = \left(  \begin{array}{cc} \frac{i \sigma_2 }{2} \tilde{T}  & \frac{1}{\sqrt{2}} Q_{(2,2)}  \\ 
- \frac{1}{\sqrt{2}} Q_{(2,2)}^T  &   \frac{i \sigma_2 }{2} \tilde{T} \end{array} \right) 
\,, \qquad  \hat \psi_1 = \left(  \begin{array}{cc} \frac{i \sigma_2 }{2} T_1  & 0  \\ 
0  &  -  \frac{i \sigma_2 }{2} T_1 \end{array} \right) 
\eeq
with  $Q_{(2,2)} =    \left(\begin{array}{cc} T & X  \\ 
B  &  T_X
\end{array} \right) $ being a bi-doublet under the $SU(2)_L \times SU(2)_R$ symmetry. Similarly to the pNGBs, the  composite partners must be rotated to   the misaligned representation $\psi_5 =  U_{\alpha}\hat \psi_5 U_{\alpha}^T$ and $\psi_1  = U_\alpha \hat \psi_1 U_\alpha^T$. Their interactions to the pNGBs and to the EW gauge bosons is determined by Eq.~\eqref{eq:Lcomposite}, where the two masses $M_A \equiv M_5$ and $M_1$ are generated by the strong dynamics alone and, in general, $M_5 \neq M_1$. 
To couple the chimera baryons to the top fields, they need to be matched to composite operators transforming as a complete representation of $\mathcal{G}$. The intuitive way to understand this is that baryons are made of confining fermions, $Q$ and $\chi$, that transform under $\mathcal{G}$; however, the same $\mathcal{H}$ representation can be embedded in different  operators. In our specific case, we have two possibilities:
\begin{itemize}
\item[a)] The baryons come from the adjoint representation, corresponding to the operator $\bar{Q} Q \bar{\chi} = (\bf{15}, \bf{\bar 6})$ of SU(4)$\times$SU(6). This can only contain $\psi_5$.
\item[b)] The baryons emerge from the antisymmetric representation, corresponding to the operator $Q Q \chi = (\bf{6}, \bf{6})$ of SU(4)$\times$SU(6). Both $\psi_5$ and $\psi_1$ match to this composite operator.
\end{itemize}

To implement partial compositeness, the SM fields for top and bottom must be included into spurions, i.e. incomplete  representations of $\mathcal{G}$. All possible spurions for the left-hand $(t_L, b_L)$ and right-hand $t_R$ fields are listed in Appendix~\ref{Appendix3}.  We recall that the SM spurions will stay in the unrotated basis, in order to preserve the EW properties of the elementary fields.

\subsection{Partial compositeness}
We can now construct the partial compositeness operators in the form of $\mathcal{G}$--invariant operators, following  the rules  illustrated in Sec.~\ref{sec:PC}. This constitutes a template model we will use to discuss general properties of partial compositeness in  CHMs. For the top partners transforming as an antisymmetric representation of $\mathcal{H}$, the SM spurions can be chosen either in adjoint  or the antisymmetric representation of $\mathcal{G}$. The two possibilities lead to inequivalent models, as follows. 
\begin{itemize}
\item[{(1)}] For the adjoint embedding of the SM $(t_L, b_L)$ and $t_R$ fields, we find two possibilities for the doublet and two for the singlet. In fact, the adjoint contains both an antisymmetric and a symmetric of Sp(4). The spurions $D_{L,A/S}$ and $D_{R,A/S}$, given in Appendix~\ref{Appendix3},  lead to the following 4 operators:
\beq
\mathcal{L}_{mix} &=&  y_{L1} f\ tr \left[ D_{L, A}^{\dag} \gamma_0 U_{\Pi}  \psi_5  \Sigma_\alpha^* U_{\Pi}^\dag \right]  +  y_{R1} f\  tr \left[ D_{R, S}^{\dag} \gamma_0 U_{\Pi}  \psi_5  \Sigma_\alpha^* U_{\Pi}^\dag \right]  \nonumber \\
&+& y_{L2}f \  tr \left[ D_{L, S}^{\dag} \gamma_0 U_{\Pi}  \psi_5  \Sigma_\alpha^* U_{\Pi}^\dag \right]  +  y_{R2}  f\ tr \left[ D_{R, A}^{\dag} \gamma_0 U_{\Pi}  \psi_5  \Sigma_\alpha^* U_{\Pi}^\dag \right] + \mbox{h.c.} \, \label{DPC}
\eeq
where  $\psi_5 \Sigma_\alpha^* $ transforms as $h \psi_5 \Sigma_\alpha^* h^{-1}$, that ensures the $\mathcal{G}$ invariance.  Here the couplings $y_{L2}$ and $y_{R2}$ break the $\eta \to -\eta$ parity and allow for mixing of the EW singlet $\tilde T$ with the top, while preserving CP--invariance. These spurions also contribute to the Higgs potential via operators in the form $tr \left[ D_{L/R}^c \Sigma^\dag D_{L/R} \Sigma \right]$ \cite{Alanne:2018wtp}, leading to the following condition to ensure the absence of a tadpole for $\eta$:
\beq
y_{L1}^* y_{L2} - y_{L2}^* y_{L1} = 0\,.
\eeq
Expanding Eq.~\eqref{DPC} gives rise to the mixing mass matrices in the top sector for $Q=2/3$ fields $(t, T, T_X, \tilde T, T_1)$ and in the bottom sector for $Q=-1/3$ fields $(b, B)$:
\beq
&& M_{2/3} =  \left( \begin{array}{ccccc}
0 & y_{L1}f c^2_{\alpha/2} & y_{L1} f s^2_{\alpha/2} & -\frac{y_{L2} f}{\sqrt{2}} s_\alpha & 0 \\
-\frac{y_{R1}f }{2} s_{\alpha} & M_5 & 0 & 0 & 0  \\
\frac{y_{R1}f }{2} s_{\alpha} & 0 & M_5 & 0  & 0  \\
y_{R2} f c_{\alpha} & 0 & 0 & M_5 & 0 \\
0 & 0& 0 & 0 & M_1
\end{array} 
\right)  
\,,  \nonumber \\  && M_{-1/3} = \left( \begin{array}{cc} 
0 & y_{L1}  c_\alpha \\
0 &  M_5
\end{array}\right) \,.
\label{AdjMass}
\eeq
an interesting pattern emerges for the top mass generation: if $t_L$ is in the antisymmetric component, $t_R$ must be in the symmetric, and vice-versa.

\item[{(2)}] For the antisymmetric embedding of the SM $(t_L, b_L)$ and $t_R$ fields, one finds one spurion for the left-handed doublet and two for the right-handed singlet,  given in Appendix~\ref{Appendix3}. Hence, the following five operators can be constructed:
\beq
\mathcal{L}_{mix} &=&  y_{L1} f\  tr \left[ A_L^{\dag} \gamma_0  U_{\Pi}  \psi_5   U_{\Pi}^T \right]  + y_{R1}  f\ tr \left[ A_R^{\dag} \gamma_0  U_{\Pi} \psi_5   U_{\Pi}^T \right]  
\nonumber \\ &+&   y_{L2} f \ tr \left[ A_L^{\dag} \gamma_0  U_{\Pi} \psi_1  U_{\Pi}^T \right]  + y_{R2} f\  tr \left[ A_R^{\dag} \gamma_0  U_{\Pi} \psi_1  U_{\Pi}^T \right]  \nonumber \\ &+&  y'_{R} f\  tr \left[ A_R^{(2)\dag} \gamma_0  U_{\Pi} \psi_5  U_{\Pi}^T \right]   + \mbox{h.c.} \,, \label{APC}
\eeq
where the coupling $y_{R}^\prime$ was not included in Ref.~\cite{Cacciapaglia:2015eqa}. From the Higgs potential, generated by operators in the form $tr \left[A_{L/R}^\dag \Sigma  A_{L/R}^c  \Sigma\right] +h.c. $ \cite{Alanne:2018wtp}, the condition of tadpole absence for $\eta$ reads:
\beq
y_{R}^{\prime*} y_{R1}^* - y_{R}^{\prime}  y_{R1} = 0\,.
\eeq
The mass matrices, in the same basis as above, are derived to be:
\beq
M_{2/3} &=&  \left( \begin{array}{ccccc}
0 & y_{L1}f c^2_{\alpha/2} & -  y_{L1} f s^2_{\alpha/2} & 0 &   \frac{y_{L2}}{\sqrt{2}} f s_\alpha  \\
-\frac{y_{R1}f }{\sqrt{2}} s_{\alpha} & M_5 & 0 & 0 & 0  \\
-\frac{y_{R1}f }{\sqrt{2}} s_{\alpha} & 0 & M_5 & 0  & 0  \\
y'_{R} f  & 0 & 0 & M_5 & 0 \\
y_{R2} f  c_{\alpha} & 0& 0 & 0 & M_1
\end{array} 
\right) 
\,,  \nonumber \\  
 M_{-1/3} &=& \left( \begin{array}{cc} 
0 & y_{L1} \\
0 &  M_5
\end{array}\right) \,.
\label{AntMass}
\eeq
For $y'_{R} =0$ and $M_5 \sim M_1$, the antisymmetric embedding will give rise to a similar mass spectrum as that of the adjoint one. 
\end{itemize}
In order to generate the top quark mass in the SU(4)/Sp(4) CHM, it is sufficient to consider a subset of the couplings that guarantee mixing of $(t_L,b_L)$ and $t_R$ with only one type of top partners, i.e. the bi-doublet or a singlet top partner. This also applies to more general CHMs with UV completion.  For example, in the SU(6)/SO(6) CHM~\cite{Cacciapaglia:2019ixa, INprep2022},  a mixing pattern can be generated by a bi-triplet  $(3, 1) \oplus (1, 3)$ under the SU(2)$_L \times$SU(2)$_R$ symmetry.  Since the simplified mixing patterns  are enough to enlighten us about the partial compositeness phenomenology, we will not consider the most general mixture scenario in this paper. We will therefore consider two subcases:
\begin{itemize}
\item[-] Bi-doublet : we can obtain this pattern by turning on only $y_{L1/R1}\neq 0$ for both adjoint and antisymmetric cases, while setting all other couplings to zero.
\item[-] Singlet : this case is obtained by setting $y_{L2/R2} \neq 0$ in both cases, with the exception of the antisymmetric where $y'_R \neq 0$ can be kept.
\end{itemize}
Assuming both  pre-Yukawa couplings  are  of the same order,  the mass matrices can be diagonalised perturbatively given that  the terms related to the EW symmetry breaking are subdominant, i.e.  $\frac{y_{R1 f}}{2} \sin \alpha \ll M_5$ and $y_{L1} \sin^2 \frac{\alpha}{2} \ll \frac{y_{R1}}{2} \sin{\alpha}$; or $\frac{y_{L2 f}}{\sqrt{2}} \sin \alpha \ll M_5$. We can  first define  the leading order (LO) rotation  in the left- and right-handed sectors as:  
\beq 
\mbox{Bi-doublet}:  \sin \phi_L  =  \frac{f y_{L1}}{\sqrt{M^2_5 + f^2 y^2_{L1}}} \,; \quad \mbox{Singlet}:  \sin \phi_R  =  \frac{f y_{R2}}{\sqrt{M^2_5 + f^2 y^2_{R2}}}  
\eeq
Then the perturbation constraints are  translated into the bound of mixing angles and mass of top partners:
\beq
\sin 2 \phi_{L/R} \gg \frac{ 2 m_t}{M_{T/\tilde{T}}} \,, ~\qquad~  \sin^2 \phi_{L} \ll \frac{m_t}{m_T} \frac{1}{\sin^2 \frac{\alpha}{2}} \,. \label{pert}
\eeq
The mass patterns can be classified for the simplified scenarios.  For the bi-doublet mixing scenario,  the masses in $(T, B)$ or $(X, T_{X})$  split at  $\mathcal{O}(f^2 \sin^2 \alpha)$ and the spectrum is:
\beq
m_t &=&  \frac{f^2 \sin \alpha  y_{L1} y_{R1}}{2 \sqrt{M_5^2 + f^2 y_{L1}^2}}  \,, \quad m_{T_X} = M_5+  \frac{f^2 \sin^2 \alpha y_{R1}^2}{8 M_5} \nonumber \\
m_T &=&  \sqrt{M^2_5 + f^2 y^2_{L1}}  + \frac{f^2 \sin^2\alpha \left(M_5^2 \left(y_{R1}^2-2 y_{L1}^2\right)-2 f^2 y_{L1}^4\right)}{8 \left(M_5^2 + f^2 y_{L1}^2\right)^{3/2}}\nonumber \\
 m_B &=&  \sqrt{M_5^2 + f^2 y_{L1}^2}-\frac{f^2 \sin^2 \alpha y_{L1}^2}{2 \sqrt{M_5^2 + f^2 y_{L1}^2}} \,, \quad m_{X} = M_5 \,, \quad   m_{T_1} = M_1
\eeq
The mass spectrum in the case of  the singlet mixing is much simpler,  and we are going to  label the one in the adjoint embedding as $D_1$ and the other option in the antisymmetric embedding as $A_1$.  For $D_1$ scenario,  the mass spectrum is:
\beq
 m_t  =  \frac{f^2 \sin \alpha y_{L2} y_{R2}}{\sqrt{2} \sqrt{f^2 y_{R2}^2+M_5^2}} \,, ~~ m_{\tilde{T}}^{(0)}  = \sqrt{M^2_5 + f^2 y^2_{R2}} \, \quad ~~ m_{Q} =  M_5 \,, ~ m_{T_1} = M_1\,;
\eeq 
while for $A_1$ scenario with $y_{R}^\prime =0$,  exchanging the role of  $\tilde{T}$ and $T_1$,  we obtain:
\beq
 m_t = - \frac{f^2 \sin \alpha y_{L2} y_{R2}}{\sqrt{2} \sqrt{f^2 y_{R2}^2+M_1^2}} \,, ~~  m_{T_1}^{(0)}  = \sqrt{M^2_1 + f^2 y^2_{R2}} \, \quad ~~ m_{Q}  =  M_5 \,, ~ m_{\tilde{T}} = M_5\,.
\eeq

\section{Electroweak Precision Test} \label{sec:ewt}

The contribution to EW precision observables (EWPO), encoded into the Peskin-Takeuchi parameters \cite{Peskin:1990zt,Peskin:1991sw}, has been widely studied in the literature, see for instance Refs~\cite{Agashe:2005dk,Grojean:2013qca,Contino:2015mha,Contino:2015gdp,Ghosh:2015wiz}. Here we present the first complete and accurate results up to order $\sin^2 \alpha$ and at one loop. The effects can be divided into three categories:
\begin{itemize}
\item[A)] Modification of the Higgs couplings and loops of other EW resonances (other pNGBs, vector and axial-vector resonances).
\item[B)] Top and top partner loops via the mixing.
\item[C)] Top partner loops via their modified gauge couplings (misalignment effect). 
\end{itemize}
While the first two are already know \cite{Contino:2015mha}, and top loops via mixing have been  computed before in the EW basis  \cite{Grojean:2013qca}. The effect of the misalignment has never been systematically discussed in the literature. As we will see, it can be dominant as it contains a logarithmic divergent term and, depending on the model, misalignment effects appear either in $S$ or in both $S$ and $T$.
This is an important new result of our work.

Firstly, the reduced Higgs couplings give rise to a well-known logarithmic contribution:
\beq
\Delta T_h &=& -\frac{3 }{8 \pi \cos^2 \theta_W} \left((1-\kappa_V^2) \log \frac{\Lambda }{m_h} + \log \frac{m_h }{m_{h,ref}}\right) \,, \nonumber \\
\Delta S_h &=& \frac{1}{6 \pi} \left((1-\kappa_V^2)\log \frac{\Lambda}{m_h} +  \log \frac{m_h }{m_{h,ref}}\right) \,, \label{TS0}
\eeq
with  $\Lambda = 4 \pi f$ being the cut-off scale of the effective CHM.  Here we will  assume that the impact of spin-1 resonance is  in the decoupling limit with $g/\tilde{g} \ll1 $ and $r \sim 1$~\cite{BuarqueFranzosi:2016ooy}, and  we can use  $\kappa_V = \cos \alpha $ in the  analysis and we neglect other resonance loops. In particular, additional pNGBs do not contribute in the minimal sector, but may give sizeable effects in larger cosets \cite{Cacciapaglia:2019ixa}.
In this work we consider the above equations as the template contribution of the EW sector of CHMs.
 
Secondly, we focus on the contribution from  composite top partners that are necessary for top partial compositeness. This includes  the mixing contribution stemming from the rotation from the gauge basis to the mass eigenstates: these terms are usually computed in the literature. The other important part is generated by the misalignment effect starting at $\mathcal{O}(\sin^2 \alpha)$, as encoded in the CCWZ objects $d_\mu$ and $E_\mu$, which modifies the couplings of the $W$ and $Z$ bosons to the top partners (C.f. the Lagrangian in Eq.(\ref{Eterm}-\ref{dterm})). In fact,  the  diagonalisation  of the mass matrices in Eqs~\eqref{AdjMass} and~\eqref{AntMass} at the zeroth order for small $\alpha$  (LO) rotates the left-handed components of the doublets $(T, B)$ and $(t_L, b_L)$ and the right-handed components of the singlets $\tilde{T}$ or $T_1$ and $t_R$.  As a consequence, gauge interactions are left invariant in absence of misalignment. Therefore, the contributions of the mixing and of the misalignment in the gauge couplings arise at the same order for small $\alpha$, and they are competitive. In particular, since the rotation matrix is unitary,  the contribution to the EWPO  from this source is finite, while the contribution  from misalignment is logarithmically divergent like the contribution of the reduced Higgs couplings. At the leading order $\mathcal{O}(\sin^2 \alpha)$, the  effects from  the rotation and misalignments  are independent in simplified  scenarios, although their  interference is  generated at the higher order and negligible for a small $\sin \alpha$. 

The splitting of EWPO sources  is associated with  the  vertex product   in the  vacuum polarization amplitude (see Appendix \ref{Appendix1}). For a  two-point $VV'$  amplitude,  we can decompose  the  product  of gauge couplings  as $g_{L/R}^V g_{L/R}^{V'} = g_{L/R }^{V(0)} g_{L/R}^{V' (0)} + \Delta_{mix}(\sin^2 \alpha) + \Delta_{mis}(\sin^2 \alpha)$ in the mass basis.  In the partial compositeness paradigm,  the zeroth order term $g_{L/R }^{V(0)} g_{L/R}^{V' (0)}$ reproduces the SM  contribution plus some rotation effect due to  the mass splitting of top partners in one $SU(2)_L$ representation. The two remaining terms  at $\mathcal{O}(\sin^2 \alpha)$ only create beyond SM  corrections to  EWPO.  For $\Delta_{mix}$,  its $\sin \alpha$ dependence  comes from the rotation matrices $\Omega_{L/R}$ or $\Omega_{L/R}^d$ in Appendix \ref{Appendix2}, while  for $\Delta_{mis}$,  the origin of   $\sin \alpha$   is purely from the gauge misalignment encoded in the Lagrangian Eq.(\ref{Eterm}-\ref{dterm}).
Following Sec.~\ref{sec:su4sp4}, we will consider simplified scenarios where only one singlet or one bi-doublet contributes to top partial compositeness. For the $SU(4)/Sp(4)$ model, the $\kappa'$ term is  zero. Thus in each scenario, there are five free parameters: $(m_T, \Delta M, \kappa ,  \sin \alpha, \sin \phi_{L/R} )$, with  the  definition $\Delta M = M_5 - M_1$.  In this parameter space, two sources of  contributions can be calculated as illustrated above.
 
\subsection{Singlet mixing scenario}
Firstly, we   consider  the singlet scenario, where either $\tilde{T}$  or $T_1$ mixes with the SM top in the $D_1$ (adjoint) or $A_1$ (antisymmetric) cases, respectively.  As the bi-doublet does not mix, custodial symmetry ensures that their masses remain degenerate. From the rotation matrix in Appendix~\ref{Appendix2}, we can see that the left-handed mixing  is generated at $\mathcal{O}(\sin \alpha)$ for the singlet case, and this yields the following contribution in the $D_1$ embedding:
\beq
\Delta T_{D,\text{mix}} &=& \frac{N_c}{16 \pi \sin^2 \theta_W \cos^2 \theta_W} \frac{\cos^2\phi_R }{\sin^2\phi_R} \frac{m_t^2}{m_{\tilde T}^2} \Big[  \left[\theta_+(y_{\tilde{T}}, y_b) 
- \theta_+(y_{\tilde{T}}, y_t) \right]- \theta_+(y_t, y_b) \Big] \,, \label{TD}
\eeq
\beq
\Delta S_{D,\text{mix}} &=& \frac{N_c}{2 \pi }\frac{\cos^2\phi_R }{\sin^2\phi_R} \frac{m_t^2}{m_{\tilde T}^2} \Big[ \psi_+(y_{\tilde{T}}, y_b) -  \chi_+( y_{\tilde{T}} , y_t) -\psi_+(y_t, y_b) \Big] \,,  \label{SD}
\eeq
\beq
\Delta U_{D,\text{mix}} &=& - \frac{N_c}{2 \pi }\frac{\cos^2\phi_R }{\sin^2\phi_R} \frac{m_t^2}{m_{\tilde T}^2} \Big[ \chi_+(y_{\tilde{T}}, y_b) -  \chi_+(y_{\tilde{T}} , y_t) -\chi_+(y_t, y_b) \Big] \,, 
\eeq
with
\beq
\psi_+ (y_{\alpha}, y_{i}) =  \frac{1}{3} \left( Q_\alpha - Q_i \right) - \frac{1}{3} (Q_\alpha + Q_i)  \log \left(\frac{y_\alpha}{y_i}\right) \,, \label{psi+}
\eeq
where  $y_i \equiv m^2_i/m^2_Z$. In Appendix~\ref{Appendix1} we provide a general result for $\psi_+(y_\alpha, y_i)$ from top partners in an irreducible representation of $SU(2)_L \times U(1)_Y$, along with  the usual  $\theta_\pm $ and $\chi_\pm $ functions.  In case of a singlet, Eq.(\ref{psi+})  matches to the  result in \cite{Lavoura:1992np}.  For the $A_1$ embedding, one simply needs to replace  $\tilde T$ with  $T_1$. Note that the effective mixing angle of  top partner  actually is $\sin \theta_{L, eff} = \frac{\cos \phi_R}{\sin \phi_R} \cdot \frac{m_t}{m_{\tilde T}}$, with the top quark mass $ m_t \propto \sin \alpha$. 

The modified gauge couplings are captured in Eq.~\eqref{eq:Lcomposite}, with the $\kappa, \kappa'$ terms  depending on the symmetry breaking pattern. In the SU(4)/Sp(4) CHM, the misalignment in the fermion sector is generated  among the  components in 4-plet $Q_{(2,2)}$ and the singlet $T_1$.   However, the LO mixing rotates in the top fields and will affect the final contribution to the oblique parameters. For the $D_1$ embedding,  this  misalignment is  independent  to the mixing sector $(t, \tilde T)$ involved in the partial compositeness and can be evaluated exactly. Using Eq.~\eqref{DAV} in the Appendix, the corresponding $\Delta S_{D,\text{mis}}$ is derived to be:
\beq
 \Delta S_{D,\text{mis}} &=& \frac{N_c \sin^2 \alpha }{2 \pi} \Big[  \left[2 -  \kappa^2  \right] (\frac{1}{3} - \frac{1}{3} \log y_{Q}^2)  + \kappa^2 \Big[(\frac{1}{3} + \frac{1}{3} \log y_{T_1}^2)  \nonumber \\ && + 2  \psi_-(y_{T_1}, y_{Q}) - 2  \chi_- (y_{T_1}, y_{Q})  - 2 \chi_+(y_{T_1}, y_{Q}) \Big]  \nonumber \\
 && + \frac{4}{3}[1 -\kappa^2] \left( \log\frac{\Lambda^2}{m^2_Z} -\frac{7}{6} \right) \Big]\,,  \label{SD1}
\eeq
where a logarithmic divergent term  emerges due to the  unitarity violation. Notice that  a  benchmark point  exists where the  divergence is cancelled between the $tr[ \bar \psi_5  d_\mu \gamma^\mu \psi_1] $ and $tr[\bar{\psi}_5 E_\mu  \gamma^\mu \psi_5] $ terms, if the coefficient  of  $tr[ \bar \psi_5  d_\mu \gamma^\mu \psi_1] $  satisfies $\kappa =1$.

For the $A_1$ embedding,  the situation is slightly different because of the mixing in  $(t, T_1)$. Neglecting the  interference at $\mathcal{O}(\sin^3 \alpha)$, this leads to the following $\Delta S_{A,\text{mis}}$:
\beq
 \Delta S_{A,\text{mis}} &=& \frac{N_c }{4 \pi} \Big[  \left[ 16 \sin^2 \frac{\alpha}{2}  - 2 \kappa^2 \sin^2 \alpha) \right] (\frac{1}{3} - \frac{1}{3} \log y_{Q}^2) \nonumber \\ && +  \kappa^2 \sin^2 \alpha \Big[  (\cos^2 \phi_R +1 )(\frac{1}{3} + \frac{1}{3} \log y_{T_1}^2) +  \sin^2 \phi_R  (\frac{1}{3} + \frac{1}{3} \log y_{t}^2)   \nonumber \\ & - &  4 \cos \phi_R  [\chi_-(y_{T_1}, y_{Q}) -   \psi_- (y_{T_1}, y_{Q}) ] - 2 (\cos^2 \phi_R +1 ) \chi_+(y_{T_1}, y_{Q})  \nonumber \\ &-& 2  \sin^2 \phi_R \chi_+ (y_t, y_{Q}) \Big]  + \frac{8}{3}[4 \sin^2 \frac{\alpha}{2}  -\kappa^2 \sin^2 \alpha]  \left( \log\frac{\Lambda^2}{m^2_Z} -\frac{7}{6} \right)  \Big]\,.  \label{SA1}
\eeq
Note that we recover Eq.~\eqref{SD1} at $\mathcal{O}(\sin^2 \alpha)$ in the limit of  $\sin \phi_R = 0$. Thus only minor differences are expected between the cases $D_1$ and $A_1$ in the large $\sin \phi_R$ region.  We would like to remark that,  since  the misalignment respects the custodial symmetry in both singlet mixing pattens,  the corresponding $\Delta T_{\rm mis}$ and $\Delta U_{\rm mis}$ vanish.


\subsection{Bi-doublet mixing scenario}
Now we turn to the mixing contribution in the bi-doublet  scenario, in such case the custodial symmetry is conserved  for  the  basis rotation till $\mathcal{O}(\sin^2 \alpha)$. The direct calculation  to this order gives:
\beq
\Delta T_{\rm mix} &=& \frac{N_c}{16 \pi \sin^2 \theta_W \cos^2 \theta_W} \Big[ \frac{\cos^2\phi_L }{\sin^2\phi_L} \frac{m_t^2}{m_T^2} \theta_+(y_t, y_B) + \frac{1}{\sin^2 2 \phi_L} \frac{4 m_t^2}{m_T^2} \theta_+(y_X, y_t)  \nonumber \\
&-& \frac{\cos^2\phi_L}{\sin^2\phi_L} \frac{m_t^2}{m_T^2} \theta_+(y_t, y_T) -  \frac{1}{\sin^2 2 \phi_L}\frac{4 m_t^2}{m_T^2} \theta_+(y_{T_X}, y_t) \Big]  \simeq   \mathcal{O} (\epsilon^4)\,,
\eeq

\beq
\Delta S_{\rm mix} &=& \frac{N_c}{2 \pi } \Big[ \frac{\cos^2\phi_L }{\sin^2\phi_L} \frac{m_t^2}{m_T^2} \bar{\psi}_+(y_t, y_T) + \frac{1}{\sin^2 2 \phi_L} \frac{4 m_t^2}{m_T^2} {\bar \psi}_+(y_{T_X}, y_t)   + 2 \bar{\psi}_+(y_T, y_B)  \nonumber \\
&+& 2 \bar{\psi}_+ (y_X, y_{T_X})  - \frac{\cos^2\phi_L }{\sin^2 \phi_L} \frac{m_t^2}{m_T^2} \chi_+(y_t, y_T) - \frac{1}{\sin^2 2 \phi_L} \frac{4m_t^2}{m_T^2} \chi_+(y_{T_X}, y_t)  \Big]   \,, \label{SBdR}
\eeq
with 
\beq
&&\bar{\psi }_+ (y_{i}, y_{j}) = \frac{2}{3}\left( Y_{L}^i - Y_{L}^j \right) - \frac{2}{3} Y^{vq} \log \left(\frac{y_i}{y_j}\right) \,, \label{Psibar+}
\eeq
where  $Y^{vq}$ is the hyper-charge of vector-like quark and  the NLO right-handed rotation is  multiplied by  $\frac{m_t}{m_T}$.  In analogy to  the case  of one  irreducible representation, Eq.(\ref{SBdR}) is transformed from the general formula Eq.~\eqref{DAV}, by defining  a new function $\bar{\psi}_+(y_i, y_j)$  for the bi-doublet  scenario as a result of  the divergence cancellation. Also we need the mass difference inside the bi-doublet till $\mathcal{O}(\sin^2 \alpha)$ for an accurate evaluation of $\Delta S$: 
\beq
m_{T} -m_{B} =  \frac{\cos^2 \phi_L}{2 \sin^2 \phi_L} \frac{m_t^2}{m_{T}^{(0)}} + \frac{\sin^2 \alpha  \sin^2 \phi_L}{4}  m_{T}^{(0)}  \,, ~~~
 m_{T_X}-m_X = \frac{2\cos \phi_L}{ \sin^2 2 \phi_L} \frac{m_t^2}{m_{T}^{(0)}} \,. 
\eeq

Differently from the singlet case, in the bi-doublet case the misalignment contributes to both $T$ and $S$.  In fact, by substituting the LO  left-handed rotation of $(t, T)$ and $(b, B)$ into  Eqs~(\ref{Eterm}-\ref{dterm}), the custodial symmetry  is  violated at $\mathcal{O}(\sin^2 \alpha)$. The corresponding  $\Delta T$ and $\Delta S$ are derived to be:
\beq
\Delta T_{\rm mis} &=& \frac{N_c \sin^2 \phi_L}{16 \pi \sin^2 \theta_W \cos^2 \theta_W} \left[ \frac{\kappa^2 \sin^2\alpha}{2} \Big[ \theta_+(y_{T_1}, y_b) - \theta_+ (y_{T_1}, y_t)\Big] -2 \sin^2 \frac{\alpha}{2}  \theta_+ (y_t, y_b)  \right.  \nonumber \\ 
&+&  \left. \left(2 \sin^2 \frac{\alpha}{2} -\frac{\kappa^2}{2}  \sin^2 \alpha  \right) \Big[ (y_t -y_b)  \left( \log\frac{\Lambda^2}{m^2_Z} -\frac{1}{2} \right) -  2 (y_t \log y_t  - y_b \log y_b ) \Big] \right] \,, \label{TBd}
\eeq
\beq
\Delta S_{\rm mis} &=& \frac{N_c}{4 \pi } \Big[ \sin^2 \phi_L \left[ \left(\frac{4}{3} \sin^2\frac{\alpha}{2} - \frac{\kappa^2}{2} \sin^2 \alpha\right)  \left(\frac{1}{3} - \frac{1}{3} \log y_t^2\right)- \frac{4}{9} \sin^2\frac{\alpha}{2} \right] \nonumber \\ &+& (\cos^2 \phi_L +1 )\left( \frac{4}{3} \sin^2\frac{\alpha}{2} - \frac{\kappa^2}{2} \sin^2 \alpha\right)  \left(\frac{1}{3} - \frac{1}{3} \log y_T^2\right) \nonumber \\
&+& \left( \frac{40}{3} \sin^2\frac{\alpha}{2} - \kappa^2 \sin^2 \alpha \right)  \left(\frac{1}{3} - \frac{1}{3} \log y_{T_X}^2\right) + 2 \kappa^2 \sin^2 \alpha \left(\frac{1}{3} + \frac{1}{3} \log y_{T_1}^2\right) \nonumber \\
&-&\kappa^2  \sin^2 \alpha \Big[ \sin^2 \phi_L \chi_+ (y_t, y_{T_1}) + (\cos^2 \phi_L +1)  \chi_+ (y_{T}, y_{T_1}) + 2 \chi_+(y_{T_X}, y_{T_1})  
 \nonumber \\ & + & 2 \cos \phi_L  \left[\chi_{-}(y_{T}, y_{T_1}) - \psi_{-}(y_{T}, y_{T_1}) \right] +  2 \left[ \chi_{-} (y_{T_X}, y_{T_1}) -  \psi_{-} (y_{T_X}, y_{T_1})  \right] \Big]  \nonumber \\ &+& \frac{8}{3} \left[ 4 \sin^2 \frac{\alpha}{2} - \kappa^2 \sin^2 \alpha \right]  \left( \log\frac{\Lambda^2}{m^2_Z} -\frac{7}{6} \right)  \Big] \,, \label{SBd}
\eeq
\beq
\Delta U_{\rm mis} &=&\frac{N_c \sin^2 \phi_L }{2 \pi}\Big[ \frac{1}{3} \left(2 \sin^2 \frac{\alpha}{2} -\frac{\kappa^2}{2}  \sin^2 \alpha  \right)   \log\frac{y_t}{y_b}  
+  2 \sin^2 \frac{\alpha}{2}  \chi_+ (y_t, y_b)  \nonumber \\ &+&  \frac{\kappa^2 \sin^2\alpha}{2}  \left[ \chi_+(y_{T_1}, y_t) - \chi_+ (y_{T_1}, y_b)\right]\Big]\,.
\eeq
Note that the first term in Eq.(\ref{TBd}) can be expanded as $\theta_+(y_{T_1}, y_b) - \theta_-(y_{T_1}, y_t) \simeq (y_t - y_b) \left( 2 \log \frac{y_{T_1}}{y_b} -3 \right)$  in the large $y_{T_1} $ limit and similar to the one in  Eq.(\ref{TD}) originating from the singlet rotation. Furthermore  other terms in Eq.(\ref{AVV}) that are proportional to $g_L^V g_R^V$ will not show up in the misalignment contribution because the LO rotation involves only left-handed fields. Concerning  the $S$ parameter,  Eq.(\ref{SBd}) behaves in a similar way like Eq.(\ref{SD1}),  with the major  difference caused by  the mixing and  mass splitting.  We  find that the logarithmic divergences  in  $\Delta T_{\rm mis}$ and $\Delta S_{\rm mis}$ simultaneously vanish at $\mathcal{O}(\sin^2 \alpha)$ for $\kappa = \frac{1}{\cos \frac{\alpha}{2}}$, while $\Delta U_{\rm mis}$ has no divergence.

\begin{figure}
	\centering 	
	\subfigure[]{\includegraphics[scale=0.8]{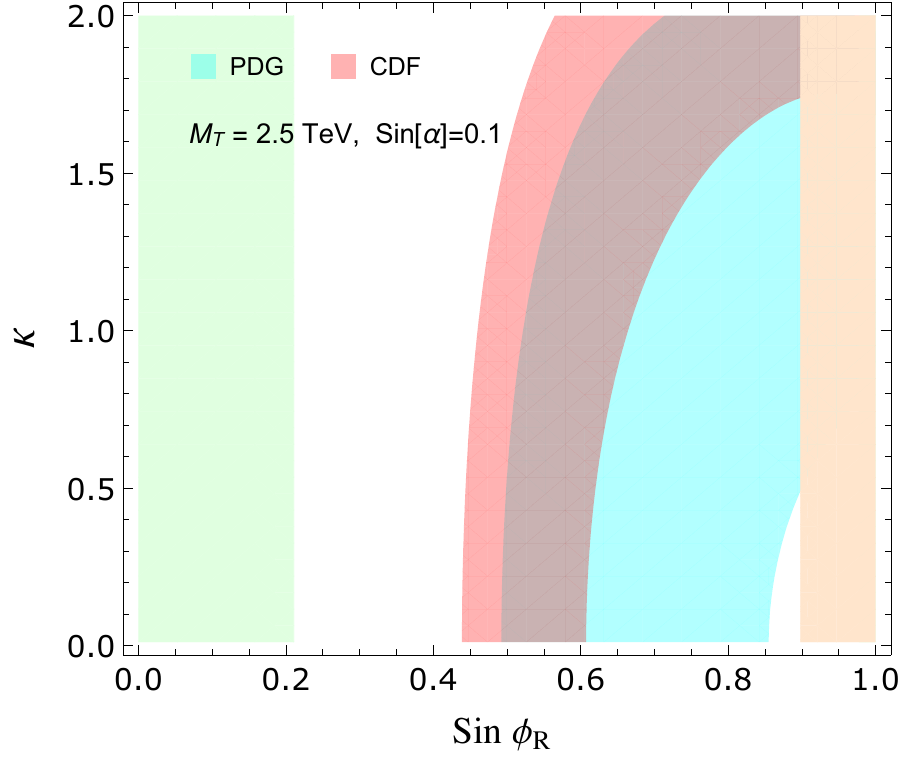}} 
	\quad	\subfigure[]{\includegraphics[scale=0.8]{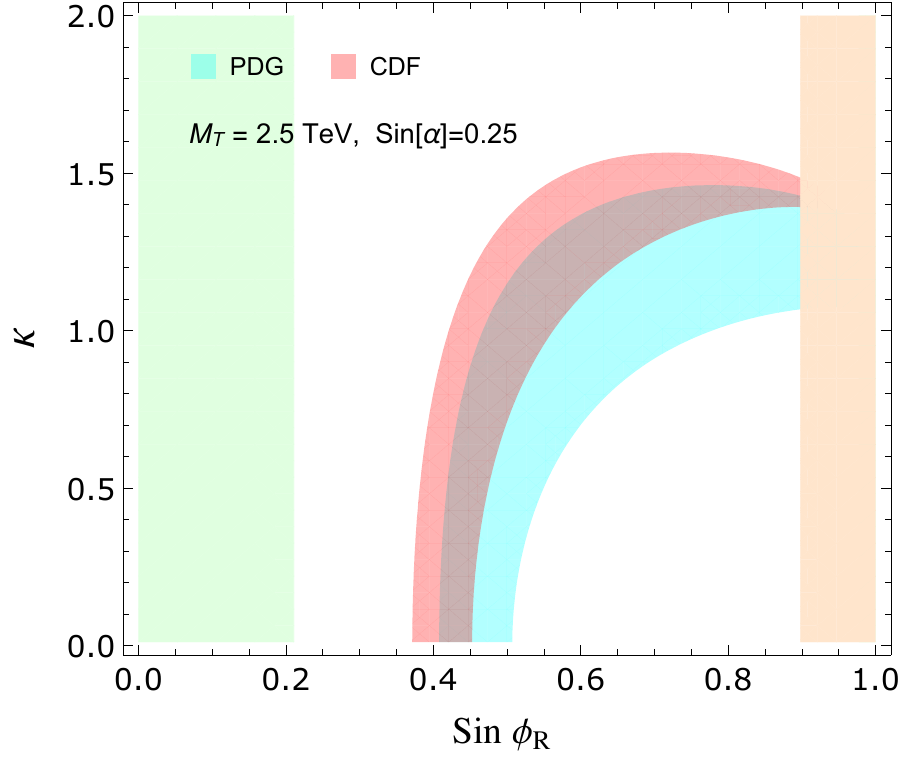}} 		
	\subfigure[]{\includegraphics[scale=0.8]{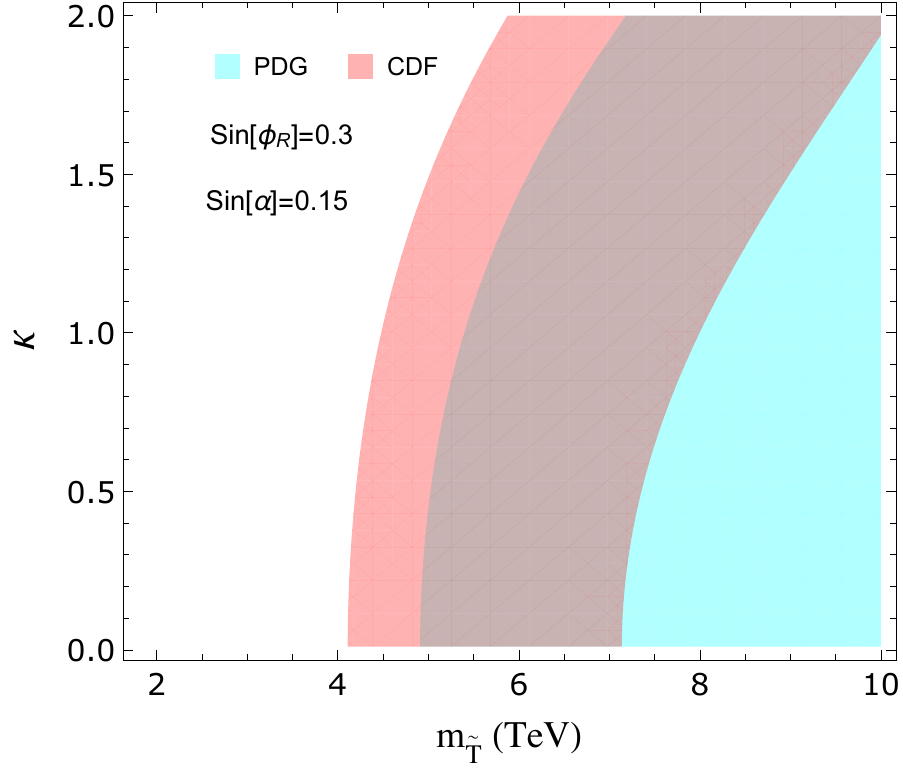}} 
	\quad \subfigure[]{\includegraphics[scale=0.8]{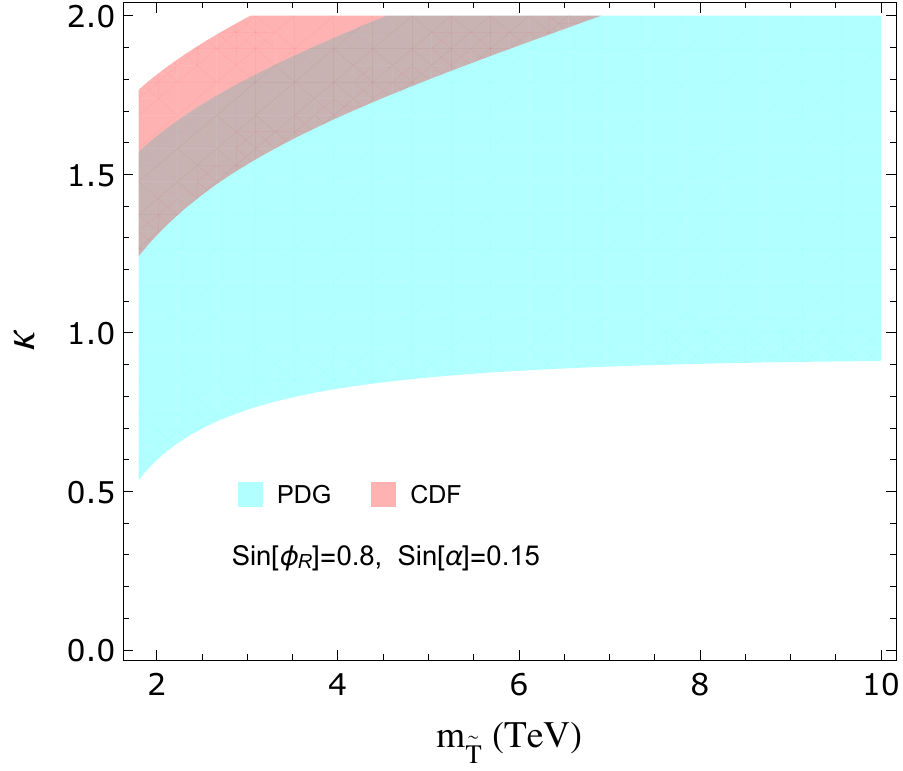}}  
	\caption{The regions satisfied  the precision measurements at  $99\%$ C.L. for  the  singlet  scenario with  $\Delta M = 0.1 $ TeV. The red band  is in light of CDF measurement of $W$ mass and the cyan region is from the PDG global fit.  In the upper panel, the green band with $\sin 2\phi_R < \frac{6 m_t}{m_{\tilde{T}}}$ indicates  our perturbation analysis breaks down,  and  the yellow band is inferred from a lower bound of  $m_{T_1} > 1.0 $ TeV \cite{ATLAS:2018ziw, ATLAS:2018alq, CMS:2018zkf}.} \label{fig:Adjsing}
\end{figure}

\begin{figure}
	\centering 	
	\subfigure[]{\includegraphics[scale=0.8]{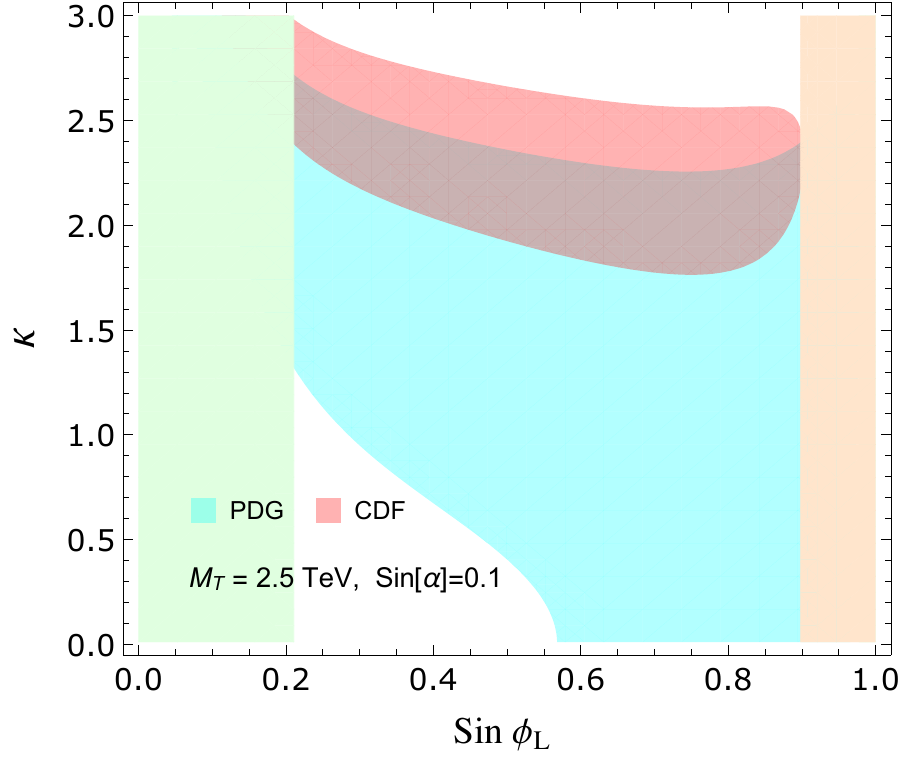}} 
	\quad
	\subfigure[] {\includegraphics[scale=0.8]{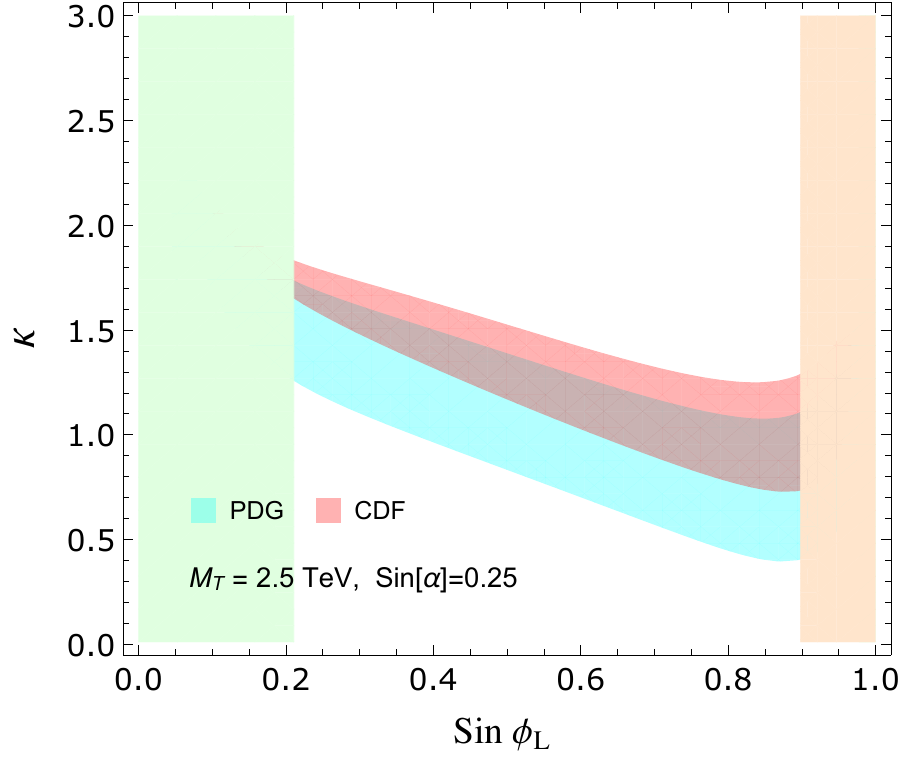}} 	\\	
	\subfigure[] {\includegraphics[scale=0.8]{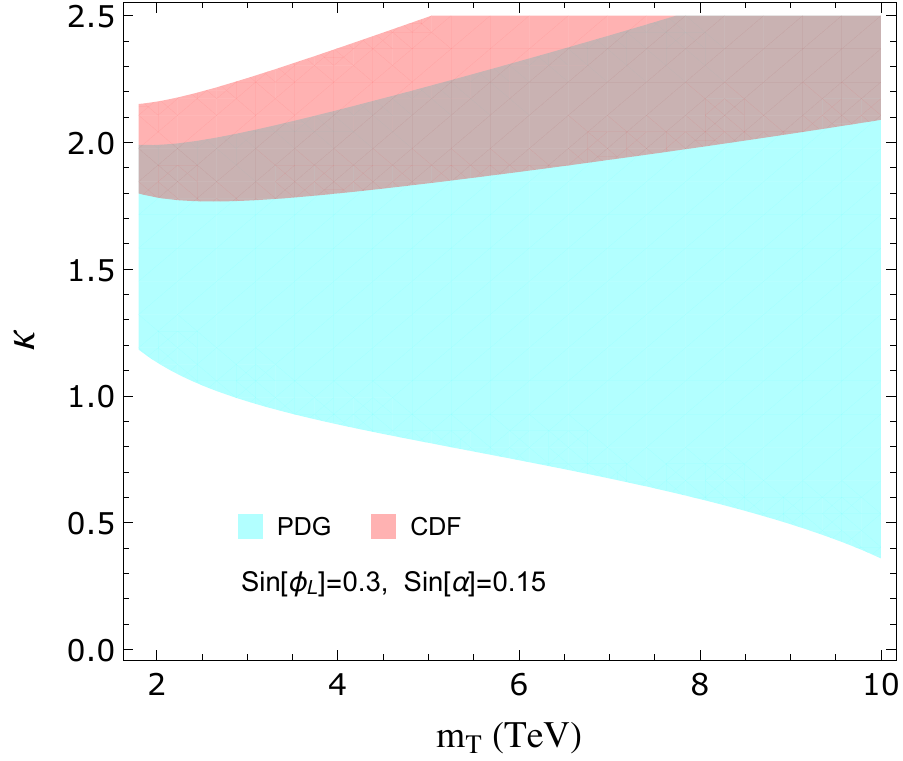}} 
	\quad \subfigure[]{\includegraphics[scale=0.8]{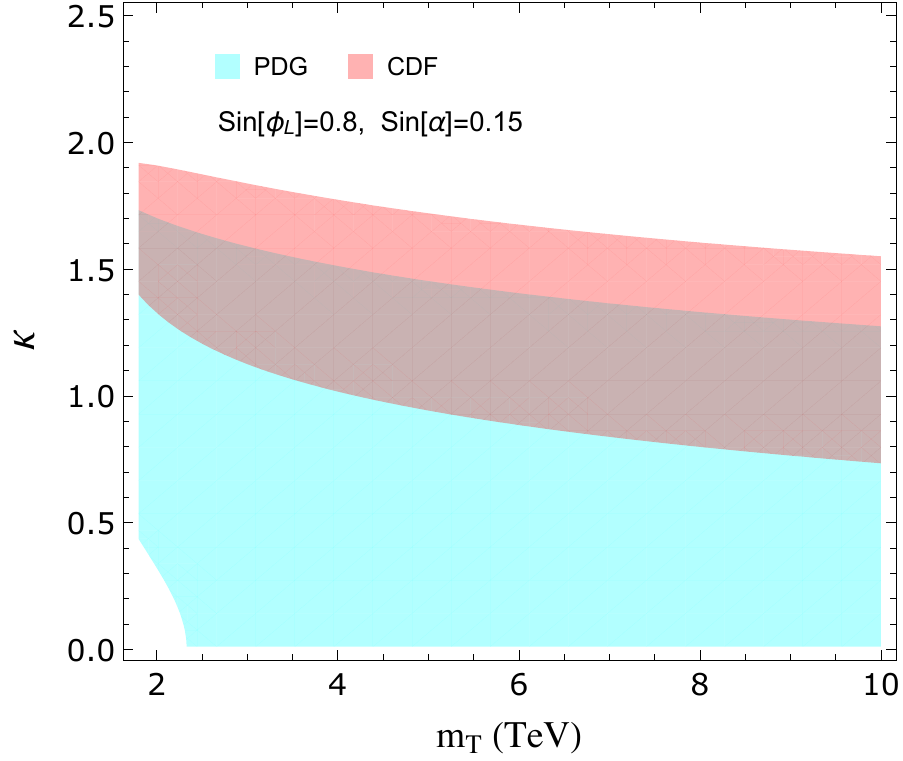}}
	\caption{The regions satisfied  the precision measurements at $99\%$ C.L.  for the bi-doublet  scenario with  $\Delta M = 0.1 $ TeV. The green and yellow bands are the same as Figure {\ref{fig:Adjsing}},  and  the  off-perturbation condition is $\sin 2\phi_L < \frac{6 m_t}{ m_T}$.}
	\label{fig:bidoub}
\end{figure}

\subsection{Analysis results}
With the analytic formulae for $S, T, U$ we can use the precision global fits to explore the allowed parameter space for the SU(4)/Sp(4) CHM.  As $U$ is always subdominant to $S$ and $T$ in our model, we can use the $U=0$  fitting contours.  The Particle Data Group (2022) \cite{Workman:2022ynf} gives the following EW global fit without the CDF result:
\beq
S = -0.01 \pm 0.07\,, \quad  T = 0.04 \pm 0.06 
\eeq
with the correlation coefficient of $0.92$.  The new CDF measurement of the W mass has been recently included in the fits: for instace in Ref.~\cite{Strumia:2022qkt} it was obtained
\beq
S = 0.06 \pm 0.08  \,, \quad T = 0.15 \pm 0.06  \label{CDF}
\eeq
with a strong correlation to be  $0.95$. We see that the central value for $T$ is significantly shifted towards positive value, while $S$ received a much less significant shift.

We will start with an order estimation to see whether a CHM is compatible with EW global fit.  Differently from the  VLQ model,   in  a CHM we will not work in the limit of  small $\sin \phi_{L/R}$  as  $M_5 \sim g_{L/R} f$.  And  due to the complexity of  gauge misalignment effect,  normally, it is  not easy to perform simplification for $S$ and $T$. While the singlet scenario is an exception,  applying  the large $M$ expansion to Eq.(\ref{TD}-\ref{SD}) gives:
 \beq
 \Delta_{T_{D, mix}} &\simeq&  \frac{3}{8 \pi \sin^2 \theta_W \cos^2 \theta_W} \frac{\cos^2 \phi_R}{\sin^2 \phi_R} \frac{y_t^2}{y_{\tilde{T}}} \left( \log \frac{y_{\tilde{T}}}{y_t} -1 \right)   \label{TD1} \\
\Delta_{S_{D, mix}} &\simeq&  \frac{1}{6 \pi}  \frac{\cos^2 \phi_R}{\sin^2 \phi_R} \frac{y_t}{y_{\tilde{T}}} \left( 2 \log \frac{y_{\tilde{T}}}{y_t}  - 5  \right)
  \eeq
 The misalignment effect in the $D_1$ scenario can also  be remarkably  simplified  if  $M_Q = M_{T_1}$ and in that degenerate limit, we have: 
  \beq 
    \Delta_{S_{D, mis}} &\simeq& \frac{2 \sin^2 \alpha}{3 \pi} (1-\kappa^2) \left( 3 \log \frac{\Lambda^2}{m_Q^2} - 2  \right)  \label{SDU}
  \eeq
Note that Eq.(\ref{TD1}-\ref{SDU})  achieve high level precision with respect to original expressions.  For  $m_{Q/\tilde{T}} \sim 2.0$ TeV,  the estimation gives $\Delta_{T_{D, mix}} \sim 0.07 \times \frac{\cos^2 \phi_R}{\sin^2 \phi_R} $  and $ \Delta_{S_{D, mis}} \sim 0.03 \times (1-\kappa^2)$, with other contributions one order smaller.   For the bi-doublet scenario,  we first simplify  $\Delta_{T_{mis}}$ by keeping the dominant terms:
\beq
\Delta_{T_{mis}} &\simeq & \frac{3  y_t \sin^2 \phi_L \sin^2 \frac{\alpha}{2} }{8 \pi \sin^2 \theta_W \cos^2 \theta_W}   \left[ (1-\kappa^2) \log \left( \frac{\Lambda^2}{m_Z^2 } \right)  + 2 \kappa^2 \log \left( \frac{y_{T_1}}{y_t} \right) \right]
\eeq
This  approximation is in lower precision  and   positive  for  $\kappa < 1$.  Analogously for $m_{T_1} \sim 2.0$ TeV,  we find  $\Delta_{T_{mis}} \sim (7 -  \kappa^2) \times 10^{-2} \sin^2 \phi_L $, and $\sin \phi_L$  is required to be large enough to compete with the negative one from $\Delta T_h$. Regarding  $\Delta S$  from top partners,  the misalignment part  is one order larger than the mixing one  if the divergence term is not suppressed. Then for  $\sin \phi_L \ll 1 $ and $M_{T/T_X} = M_{T_1}$,   $\Delta_{S_{mis}}$  in the bi-doublet scenario just reduces to Eq.(\ref{SDU}).  But  for $m_{T} \sim 2.0$ TeV and  $\sin \phi_L \sim 0.8$ (index of mass splitting), the precise value  from Eq.(\ref{SBd}) is  enhanced by around $ 25 \%$ in magnitude compared with using $m_{T}$ in Eq.(\ref{SDU}). Hence the rough evaluation indicates that within $3 \sigma$ C.L. both the singlet and  bi-doublet scenarios can  accommodate the positive shifts in CDF measurement.  In the following, we will deliver a two-parameter $\chi^2$ analysis, by inputting  the accurate  $S, T$ expressions.

In Figure \ref{fig:Adjsing},  we display the constraint of  precision measurement on the  parameter space for the  $D_1$ singlet scenario, with the 99\% C.L. region permitted by PDG (2022) shown by the cyan band, and the region including CDF by the red one.  Since all the contours are insensitive to the mass difference of $M_5- M_1$, we  set $\Delta M = 100$ GeV in the plots.  Provided that the singlet top partner $\tilde{T}$ is not too heavy,  the positive contribution in Eq.~\eqref{TD} can overcome the small negative one from the reduced Higgs coupling.  However,  $\Delta T_{\rm mix}$ in Eq.~\eqref{TD} is inversely proportional to $\sin^2 \phi_R$ in both $D_1$ or $A_1$ embeddings,  thus for a small $\sin \phi_R$  the overall positive correction is too big to fit into the electroweak precision data. This characteristic  behaviour is manifested in Figure \ref{fig:Adjsing}(a) and \ref{fig:Adjsing}(b). Also the logarithmic divergence in $\Delta S_{D(A),\text{mis}}$ will turn into dominantly negative as the value of $\kappa$ increases, this  will  translate into  an upper  bound for  the  $\kappa$ coefficient as shown in Figure \ref{fig:Adjsing}(a)- \ref{fig:Adjsing}(d).  We can  find out the $\kappa$ is mainly constrained to be  $\mathcal{O}(1)$  with  the CDF global fit preferring a larger $\kappa$  in the same set of other inputs.  Notice that the shape of the allowed region  changes dramatically  between Figure \ref{fig:Adjsing}(c) and  Figure \ref{fig:Adjsing}(d), which shows that enhancing the LO mixing $\sin \phi_R$ can alleviate the lower bound of  $m_{T}$. It should be noted that top partner masses in the multi-TeV scale, as preferred by Lattice data, can well fit within the EW precision limits.

For comparison, the  constraints from the EW global fit   is also  imposed on the bi-doublet scenario, as shown in  Figure \ref{fig:bidoub}. We find that the pattern of the plots is mainly determined by the $T$ parameter, while the $S$ parameter simply drags  $\kappa$ toward smaller values due to the logarithmic term. More in detail, we see that the first term in $\Delta T_{\rm mis}$ in Eq.~\eqref{TBd} is always positive for a TeV scale top partner and  plays the same role as  the singlet mixing contribution. Instead, the term of the second line in Eq.~\eqref{TBd} (including  the logarithmic and finite terms) will  transit from positive to negative at the point of $\kappa = \frac{1}{\cos \frac{\alpha}{2}}$. Thus, $\Delta T_{\rm mis}$ in Eq.~\eqref{TBd}, stemming from the misalignment, is positive for relatively small $\kappa$ and large $m_{T_1}$.  In contrast to the singlet case in Eq.~\eqref{TD},  the  coefficient of the  positive term  is proportional to $\sin^2 \phi_L$.   Therefore, in order to compete with the negative contribution from the Higgs couplings in Eq.~\eqref{TS0}, a minimum value of  $\sin \phi_L$ is determined either by the  $T$ parameter lower bound or by the perturbation constraint in Eq.~\eqref{pert}.  For large $\sin \phi_L$,  However, due to a reduction in $m_{T_1} = m_T \sqrt{1- \sin^2 \phi_L} - \Delta M$,  the first positive term in Eq.~\eqref{TBd} can become smaller  than the negative logarithmic term for $\kappa > \frac{1}{\cos \frac{\alpha}{2}}$. This will lead to the exclusion of some  upper region in Figure \ref{fig:bidoub}(a).  The same pattern is exhibited in  Figure \ref{fig:bidoub}(b), where the allowed region shrinks due to  a larger $\sin \alpha$.  Figure \ref{fig:bidoub}(c)  shows  the  allowed region  in the $(m_T, \kappa)$ space, where for   $\sin \phi_L  = 0.3 $, the $T$ parameter is only positive for a large $m_T$, while in Figure \ref{fig:bidoub}(d) with $\sin \phi_L =0.8$, the allowed $\kappa$ value is significantly lowered.

\section{Discussion and conclusions}  \label{sec:concl}

Composite Higgs models with top partial compositeness offer a valid alternative to the Higgs sector of the SM, as they solve the hierarchy problem in the Higgs mass by replacing an elementary scalar field by a composite state of fermions. In the recent years, this possibility has risen to the level of one of the main models for new physics to be searched for at collider experiments. Besides direct searches for resonances in the electroweak and top sectors, this class of models predicts distinctive modifications to electroweak precision observables. 

In this work, after reviewing  some universal property in CHM, we revisited the contribution of top partners to electroweak precision via the oblique $S$ and $T$ parameters. We identified a new contribution stemming from misalignment effects that modify the gauge couplings of the top partners to the $W$ and $Z$ bosons. Contrary to the usual mixing effects, these contributions can be logarithmically divergent and hence numerically large. For concreteness, we computed them explicitly in the minimal model with an underlying gauge-fermion description, based on the coset SU(4)/Sp(4). Nevertheless, we only consider simplified mixing patterns involving either a singlet or a custodial bi-doublet, henceforth the results can be generalised to other cosets in this limit. The singlet case can naturally accommodate a positive shift on $T$. For the bi-doublet case, where a negative $T$ from reduced Higgs coupling is not compensated by the basis rotation, but the misalignment effects can shift the $T$ parameter towards positive value, as suggested by the recent CDF measurement of the $W$ mass. In this case, an order unity derivative coupling  of the pNGBs to the top partners is required.

In general, we see that the misalignment contribution is crucial for a correct estimation of the impact of top partial compositeness on electroweak precision. In particular, the effect of the derivative couplings can dominate and push the $T$ parameters towards positive values. Furthermore, masses for the top partners in the multi-TeV are compatible with the new CDF measurement, which requires a sizeable positive shift in $T$. This scenario will be tested in future colliders via precision measurements at the $e+e-$ run and direct searches for top partners at the 100 TeV hadronic run.

\section*{Acknowledgments}
H.C. \ is supported by the National Research Foundation of Korea (NRF) grant funded by the Korea government (MEST) (No. NRF-2021R1A2C1005615).

\newpage
\appendix

\section{ Oblique parameters: $S, T$ and $U$} \label{Appendix1}   

In this section, we will show how to derive the  vacuum polarization amplitude  with VLQ running in the loop that are used in  the oblique parameters. Let us denote the up and down-type quarks as $\psi^{u}$, $\psi^d$  and   fermions of  exotic charges $Q = \frac{5}{3}$ as $X_{u}$ or $Q= -\frac{4}{3}$ as $X_{d}$.   The Lagrangian of $\psi^{u,d}$ coupling to  a gauge boson $W_{\mu}^{\pm, 3}, B_\mu $ in the mass basis reads:
\beq
 \mathcal{L}_{1}  &=&  \frac{g_2}{\sqrt{2}}   \left[W_{\mu}^+  \bar{\psi}^u_\alpha  \gamma^\mu \left( V_{L, \alpha i}  P_L + V_{R, \alpha i } P_R \right) \psi^d_i 
\right] + h.c.  \nonumber\\
&+& \frac{g_2}{2} W_{3 \mu} \left[ \bar{\psi}_\alpha^u  \gamma^\mu \left( U_{L, \alpha \beta}  P_L + U_{R, 
\alpha \beta} P_R \right) \psi^u_\beta -   \bar{\psi^d_i}   \gamma^\mu \left( D_{L, ij} P_L + D_{R, ij}P_R \right) \psi^d_j \right] \nonumber \\
&+& g_1 B_{\mu }  \left[ \bar{\psi}_\alpha^u  \gamma^\mu \left( Y_{L, \alpha \beta} ^u P_L + Y_{R, \alpha \beta}^u P_R \right) \psi^u_\beta + \bar{\psi}^d_i   \gamma^\mu \left( Y_{L, ij}^d P_L + Y_{R, ij}^d  P_R \right) \psi^d_j \right] 
\eeq
with $P_{L/R} =\frac{1}{2}(1\mp \gamma_5)$. 
Since the VLQ with  exotic charges will not mix with SM fermions,  their gauge interactions are:
\beq
\mathcal{L}_{2} &= &\frac{g_2}{\sqrt{2}}  W_{\mu}^+   \left[ \bar{X}_{u}  \gamma^\mu \left( \tilde{V}_{L, \alpha}  P_L + \tilde{V}_{R, \alpha } P_R \right) \psi^u_\alpha +  \bar{\psi}^d_i  \gamma^\mu \left( \hat V_{L, i}  P_L + \hat V_{R,  i } P_R \right) X_{d} \right]   \nonumber \\
&+ & \frac{g_2}{2} W_{3 \mu} \left[ \bar{X}_{u}  \gamma^\mu  U_{X}  X_{u}  -  \bar{X}_{d}    \gamma^\mu D_{X} X_{d}\right]  \nonumber \\ &+&  \frac{g_1}{2} B_{\mu} \left[ \bar{X}_{u}  \gamma^\mu (2 Q - U_{X} ) X_{u} -   \bar{X}_{d}   \gamma^\mu  (2 Q + D_{X} ) X_{d} \right] + h.c. 
\eeq
One can derive the transverse part of vacuum polarization amplitude  to be:
\beq
A_{VV} (p^2 =0 ) &=& \frac{N_c m^2_Z}{128 \pi^2}  \left[ (|g_L^V|^2  + |g_R^V|^2) \left(A_0 (m^2_1) + A_0 (m^2_2) + (m^2_1 + m^2_2) B_0 (0, m^2_1, m^2_2) \right) \right. \nonumber \\
& & \quad \qquad  - \left. 8 \, {\mathcal Re}(g_L^V g_R^{V  \dag}) m_1 m_2 B_0 (0, m^2_1, m^2_2) \right] \,, \label{AVV0}
\eeq
with the masses of two fermions  to be $m_{1,2}$  and the superscript $V$ standing for a specific gauge boson.  Note that  $g_{L/R}^V$ equals one  coupling in   $( V_{L/R},  U_{L/R}, - D_{L/R},  2 Y_{L/R}^{u,d})$ or  $(\tilde{V}_{L/R}, \hat{V}_{L/R}, U_{X}, - D_{X},  2 Y_{X_{u,d}} )$. The $A_0(m^2)$ and $B_0(p^2, m_1^2, m_2^2)$ are rescaled Veltman  scalar loop functions~\cite{tHooft:1978jhc, Passarino:1978jh}. For $p^2 =0$,  the dimensional regularization gives:
\beq
A_0(m^2) = m^2  (\Delta +1 - \log \frac{m^2}{\mu^2}) \,, 
\quad 
B_0(0, m^2_1, m^2_2) = \frac{A_0(m^2_1) -A_0(m^2_2)}{m^2_1 -m^2_2} \, \label{B0d}
\eeq
with $\Delta = \frac{2}{4-d} + \log 4\pi + \gamma_E$. The pole at $d = 4$ represents the logarithmic divergence. Using the momentum cut-off regularization, one can  calculate:
\beq
B_0 (0, m^2, m^2) = \frac{(2 \pi \mu)^{4-d}}{i \pi^2} \int \frac{ d^4 k}{(k^2 - m^2)^2} = \log \frac{\Lambda^2}{m^2} -1 \label{B0m}
\eeq
directly matching Eq.(\ref{B0d}) with Eq.(\ref{B0m}) we get: 
\beq
\Delta + \log \mu^2 + 1 = \log \Lambda^2 \,, \label{Ld}
\eeq
By substituting Eq.(\ref{B0d}, \ref{Ld}) into Eq.(\ref{AVV0}),  we can simplify the expression to be:
\beq
A_{VV}(p^2=0) &=& \frac{N_c m^2_Z}{128 \pi^2} \Big[ (|g_L^V|^2  + |g_R^V|^2) \, \theta_+ (y_1, y_2) + 2 \, {\mathcal Re}(g_L^V g_R^{V  \dag}) \,  \theta_-(y_1, y_2) \nonumber \\ &+&   (|g_L^V|^2  + |g_R^V|^2) \Big[  (y_1 +y_2) \left( \log\frac{\Lambda^2}{m^2_Z} -\frac{1}{2}\right)  - 2  \left( y_1 \log y_1 + y_2 \log y_2 \right) \Big] \nonumber \\ &+&  4 \, {\mathcal Re}(g_L^V g_R^{V  \dag}) \sqrt{y_1 y_2} \Big[  \log (y_1 y_2) - 2  \left( \log\frac{\Lambda^2}{m^2_Z} -1\right) \Big] \Big]  \label{AVV}
\eeq
with $y_i = m_i^2/m_Z^2$ being dimensionless and $\theta_\pm$ defined to be~\cite{Lavoura:1992np}:
\beq
\theta_+(y_1, y_2) &=& y_1+y_2 -\frac{2 y_1y_2 \log \left(\frac{y_1}{y_2}\right)}{y_1-y_2}   \\
\theta_-(y_1, y_2) &=& 2 \sqrt{y_1y_2} \left(\frac{(y_1+ y_2) \log
\left(\frac{y_1}{y_2}\right)}{y_1-y_2}-2\right)
\eeq
Because the term proportional to $|g_L^V|^2  + |g_R^V|^2$ gets additional contribution from $A_0$ functions compared with the one proportional to ${\mathcal Re}(g_L^V g_R^{V  \dag})$,  two divergence structures are different. Then we can calculate the derivative of $A_{VV'}$ to be:
\beq
\frac{\partial A_{VV'}}{\partial p^2}|_{p^2 =0} &=& \frac{N_c}{96 \pi^2} \left[ (g_L^V g_L^{V' \dag} + g_R^V g_R^{V' \dag}) \Big[ \frac{(m^2_1-m^2_2)^2}{4} \, \frac{\partial^2 B_0 (p^2, m^2_1, m^2_2)}{\partial^2 p^2}|_{p^2 =0}  \right. \nonumber \\
&+& \left.  \frac{ m^2_1+ m^2_2 }{2}\, \frac{\partial B_0 (p^2, m^2_1, m^2_2)}{\partial p^2}|_{p^2 =0} - B_0(0, m^2_1, m^2_2) \Big] \right. \nonumber \\ 
&-&  \left. 3 (g_L^V g_R^{V' \dag} + g_R^V g_L^{V' \dag})  m_1 m_2 \, \frac{\partial B_0 (p^2, m^2_1, m^2_2)}{\partial p^2}|_{p^2 =0}  \right]
\eeq
using the dimensional regularization,  the two-point loop function $B_0$ takes the form:
\beq
B_0 (p^2, m^2_1, m^2_2) = \Delta - \int^1_0 d x \log \left[ \frac{m^2_1 x + m^2_2 (1-x) - p^2 x (1-x)}{\mu^2} - i \varepsilon \right]
\eeq
The derivatives of $B_0$ are all  finite and can be evaluated analytically to be:
\beq
 \frac{\partial B_0 (p^2, m^2_1, m^2_2)}{\partial p^2}|_{p^2 =0}  &=& \frac{m_1^4 - m_2^4 - 2 m_1^2 m_2^2 \log \frac{m_1^2}{m_2^2}}{2 \left(m_1^2-m_2^2\right)^3} 
\label{dB0} \\
 \frac{\partial^2 B_0 (p^2, m^2_1, m^2_2)}{\partial^2 p^2}|_{p^2 =0} &=& \frac{(m^2_1 + m^2_2)^2 + 8 m^2_1 m^2_2}{3(m^2_1 - m^2_2)^4}  \nonumber \\  & - &    \frac{ 2 m^2_1 m_2^2 \left(m_1^2+m_2^2\right)  \log \left(\frac{m_1^2}{m_2^2}\right)}{\left(m_1^2-m_2^2\right)^5}  \, \label{dB1}
\eeq
In fact this approach avoids the integration complexity confronted in the average slope method  adopted by\cite{Lavoura:1992np}.  Now inserting Eq.(\ref{B0d}, \ref{dB0}-\ref{dB1}), we obtain  
\beq
\frac{\partial A_{VV'}}{\partial p^2}|_{p^2 =0} &=&  - \frac{N_c}{64 \pi^2} \Big[ (g_L^V g_L^{V' \dag} + g_R^V g_R^{V' \dag}) \left(\chi_+ (y_1, y_2 ) - \frac{1}{3} \log (y_1 y_2) +  \frac{2}{3} \left( \log \frac{\Lambda^2}{m_Z^2} -\frac{7}{6} \right) \right)  \nonumber \\ 
&+& (g_L^V g_R^{V' \dag} + g_R^V g_L^{V' \dag}) \Big(\chi_-(y_1, y_2) - \psi_-(y_1,y_2) \Big) \Big]  \,, \label{DAV}
\eeq
Note that our result only differs  from \cite{Lavoura:1992np} in the third term of the first line.  One will  get a wrong constant term $\frac{G_V+ G_A}{24 \pi^2}$ from Eq.(14) in \cite{Lavoura:1992np}, by taking the limit of $\frac{y_i}{\epsilon} |_{\epsilon\to 0}$ for conversion into derivative. The definitions of $\chi_\pm$ and  $\psi_-$ are:
\beq
\chi_+ (y_1, y_2) &=& \frac{\left(3 y_1y_2 (y_1+ y_2) - y_1^3 - y_2^3\right) \log
   \left(\frac{y_1}{y_2}\right)}{3 (y_1- y_2)^3}+\frac{5 \left(y_1^2+ y_2^2\right)-22 y_1 y_2}{9 (y_1-y_2)^2} \\
  \chi_-(y_1, y_2) &=& -\sqrt{y_1y_2} \left(\frac{y_1+ y_2}{6 y_1 y_2}-\frac{y_1+ y_2}{(y_1- y_2)^2}+\frac{2 y_1 y_2 \log \left(\frac{y_1}{y_2}\right)}{(y_1-y_2)^3}\right) \\
  \psi_-(y_1, y_2) &=& - \frac{y_1 + y_2}{6 \sqrt{y_1 y_2}}
\eeq
With the formulas of  Eq.(\ref{AVV}, \ref{DAV}), the oblique parameters $S, T, U$ can be calculated straightforward in a generic model and the divergence  will only show up when the unitarity is violated, e.g.  by the misalignment in CHM discussed in the main text.  As an application, we consider a simplified case without divergence, where VLQ is embedded in an irreducible representation of $SU(2)_L \times U(1)_Y$ and interplays with SM fermions  via a Higgs VEV insertion.  
From Eq.(\ref{DAV}), the $S$ parameter is computed by summing over all  fermion  contribution:
\beq
 S &=& -16 \pi \sum_{\{\psi, X\}} \partial A_{3Y}/\partial p^2 |_{p^2 =0}  \nonumber \\
 &=&   \frac{N_c}{2 \pi} \Big[ \sum_{ij}  {\mathcal Re}(D_{L, ij} D_{R, ij}^*)  \psi_-(y_i,y_j) + \sum_{\alpha \beta}{\mathcal Re}(U_{L, \alpha \beta} U_{R, \alpha \beta}^*)  \psi_-(y_\alpha,y_\beta) \nonumber \\
&- &  \sum_{\alpha < \beta} \Big[ \left( |U_{L, \alpha \beta}|^2  + |U_{R, \alpha \beta}|^2 \right) \chi_+ (y_\alpha, y_\beta ) + 2  {\mathcal Re}(U_{L, \alpha \beta} U_{R, \alpha \beta}^*) \chi_-(y_\alpha, y_\beta)  \Big] \nonumber \\
&- & \sum_{i<j} \Big[ \left( |D_{L, ij}|^2  + |D_{R, ij}|^2 \right) \chi_+ (y_i, y_j ) + 2  {\mathcal Re}(D_{L, ij} D_{R, ij}^*) \chi_-(y_i, y_j)  \Big]  \nonumber \\
&- &  \frac{1}{3} \Big[\sum_{\alpha} Q_\alpha \left( U_{L, \alpha \alpha}  + U_{R, \alpha \alpha} \right)  \Big(\log (y_\alpha^2 ) - 1 \Big)  -  \sum_{i}  Q_i \left( D_{L, ii}  + D_{R, ii} \right) \Big(\log (y_i^2 ) - 1 \Big) \Big] \nonumber \\
&+ &  \frac{1}{6} \Big[ \sum_{\alpha \beta}  \left( |U_{L, \alpha \beta}|^2  + |U_{R, \alpha \beta}|^2 \right) \log(y_\alpha y_\beta) + \sum_{ij}  \left( |D_{L, ij}|^2  + |D_{R, ij}|^2 \right) \log(y_i y_j) \Big] \nonumber \\
&-&  \frac{1}{3} \Big[ U_X (2Q_{X_u}- U_X) \Big(\log (y_{X_u}^2 ) - 1 \Big)  - D_X (2Q_{X_d} + D_X) \Big(\log (y_{X_d}^2) - 1 \Big) \Big]  \Big] \label{ES0}
\eeq
Using the condition that  terms proportional to $(y_a \log y_a + y_b \log y_b)$ along with the divergence $(y_a + y_b) \left(\log\frac{ \Lambda^2}{m_Z^2} -\frac{1}{2} \right)$ in the expression of $T$ parameter will vanish, we can derive that:
\beq
&& \sum_i  \left(|V_{L, \alpha i }|^2+ |V_{R, \alpha i }|^2 \right)  + \left(\tilde{V}_{L, \alpha}^2 + \tilde{V}_{R, \alpha}^2 \right) = \sum_\beta \left(|U_{L, \alpha \beta}|^2 + |U_{R, \alpha \beta}|^2 \right)  \label{sum1} \\
&& \sum_\alpha  \left(|V_{L, \alpha i }|^2+ |V_{R, \alpha i }|^2 \right)  + \left(\hat{V}_{L, i}^2 + \hat{V}_{R, i}^2 \right) = \sum_j \left(|D_{L, ij }|^2 + |D_{R, ij}|^2 \right) \label{sum2} \\
&& \sum_\alpha  \left(\tilde{V}_{L, \alpha}^2 + \tilde{V}_{R, \alpha}^2 \right) = 2 U_X^2 \,,\quad \sum_i  \left(\hat{V}_{L, i}^2 + \hat{V}_{R, i}^2 \right) = 2 D_X^2 
\label{sum3}
\eeq
Similarly in the $T$ parameter,  Requiring that terms proportional to $\sqrt{y_a y_b} \log(y_a  y_b)$ to be cancelled  will lead to:
\beq
&& \sum_i {\mathcal Re} (V_{L,\alpha i} V_{R, \alpha i}^*)  \sqrt{y_{i}} + {\mathcal Re} (\tilde{V}_{L, \alpha} \tilde{V}_{R, \alpha}^*) \sqrt{y_{X_u}} = \sum_\beta {\mathcal Re} (U_{L, \alpha \beta} U_{R, \alpha \beta}^*) \sqrt{y_\beta} \label{sum4}  \\
&& \sum_\alpha {\mathcal Re} (V_{L,\alpha i} V_{R, \alpha i}^*)  \sqrt{y_{\alpha}} + {\mathcal Re} (\hat{V}_{L, i} \hat{V}_{R, i}^*) \sqrt{y_{X_d}} = \sum_j {\mathcal Re} (D_{L,ij} D_{R, ij }^*) \sqrt{y_j}  \label{sum5} \\
&& \sum_\alpha {\mathcal Re} (\tilde{V}_{L, \alpha}  \tilde{V}_{R, \alpha}^*) \sqrt{y_{\alpha}} = U_X^2 \sqrt{y_{X_u}} \,, \quad  \sum_i {\mathcal Re} (\hat{V}_{L, i}  \hat{V}_{R, i}^*) \sqrt{y_{i}} = D_X^2 \sqrt{y_{X_d}}  \label{sum6}
\eeq 
 that are  general results for  VLQ in an irreducible representation, and we have explicitly verified that these sum rules  are observed in the non-standard doublet ($Y = \frac{7}{6}, -\frac{5}{6}$) and triplet scenarios ($Y = \frac{2}{3}, -\frac{1}{3}$). This class of models also satisfy the following relations: 
\beq
 U_{L/R} =  V_{L/R} V_{L/R}^\dag -  \tilde{V}_{L/R}^\dag \tilde{V}_{L/R} \,, \quad  && D_{L/R} =  V_{L/R}^\dag V_{L/R} - \hat{V}_{L/R} \hat{V}_{L/R}^\dag 
 \nonumber \\
 U_{X} =  \tilde{V}_{L/R} \tilde{V}_{L/R}^\dag \,, \quad  &&  D_{X}  =    \hat{V}_{L/R}^\dag \hat{V}_{L/R}  \label{UD} 
\eeq
Then applying Eq.(\ref{UD}) and Eq.(\ref{sum1}-\ref{sum3}), we can rearrange most of the terms other than $\chi_\pm$ and $\psi_-$ in  Eq.(\ref{ES0}) into a new function $\psi_+$ that is defined as:
\beq
\psi_+ (y_a, y_b) =  \frac{1}{3} \left( Q_a - Q_b \right) - \frac{1}{3} \left( Q_a + Q_b \right) \log \left(\frac{y_a}{y_b}\right)  \label{Psi+} 
\eeq
Note that $\psi_+$ is antisymmetric for exchanging $(y_a, y_b)$  and in fact $Q_a - Q_b =\pm 1$ always holds due to the $W^\pm_\mu$ current. The  structure inside the logarithm is expected since the divergence in  $S$  parameter  vanishes due to unitarity conservation.  Also using   Eq.(\ref{sum4}-\ref{sum6}), we can conduct a transformation for the  expression: 
\beq
 && \sum_{ij}  {\mathcal Re}(D_{L, ij} D_{R, ij}^*)  \psi_-(y_i,y_j) + \sum_{\alpha \beta}{\mathcal Re}(U_{L, \alpha \beta} U_{R, \alpha \beta}^*)  \psi_-(y_\alpha,y_\beta)  - \frac{1}{3} \left(U_X^2 + D_X^2 \right) \nonumber \\
 &=& 2 \sum_{\alpha i} {\mathcal Re} (V_{L,\alpha i} V_{R, \alpha i}^*) \psi_-(y_\alpha, y_i)  + 2 \sum_\alpha  {\mathcal Re}(\tilde{V}_{L, \alpha} \tilde{V}_{R,  \alpha}^*)  \psi_-(y_{X_u} ,y_\alpha) \nonumber \\ &+&  2 \sum_i {\mathcal Re}(\hat V_{L, i} \hat V_{R, i }^*)  \psi_-(y_i ,y_{X_d}) 
\eeq
Thus the general formula  for $S$ parameter originating from mixing with VLQ in an irreducible representation is:
\beq
 S  &=&   \frac{N_c}{2 \pi} \Big[ \sum_{\alpha i} \left[ \left( |V_{L, \alpha i}|^2  + |V_{R, \alpha i}|^2 \right) \psi_+ (y_\alpha, y_i) + 2 {\mathcal Re } (V_{L,\alpha i} V_{R, \alpha i}^*) \psi_-(y_\alpha, y_i) \right] \nonumber \\
 &+&  \sum_{\alpha} \left[ \left( |\tilde{V}_{L, \alpha}|^2  + |\tilde{V}_{R, \alpha}|^2 \right) \psi_+ (y_{X_u}, y_\alpha)  + 2  {\mathcal Re}(\tilde{V}_{L, \alpha} \tilde{V}_{R,  \alpha}^*)  \psi_-(y_{X_u} ,y_\alpha) \right] \nonumber \\ &+& \sum_{i} \left[ \left( |\hat{V}_{L, i}|^2  + |\hat{V}_{R, i}|^2 \right) \psi_+ (y_i, y_{X_d}) +  2 {\mathcal Re}(\hat V_{L, i} \hat V_{R, i }^*)  \psi_-(y_i ,y_{X_d}) \right] \nonumber \\
&- &  \sum_{\alpha < \beta} \left[ \left( |U_{L, \alpha \beta}|^2  + |U_{R, \alpha \beta}|^2 \right) \chi_+ (y_\alpha, y_\beta ) + 2  {\mathcal Re}(U_{L, \alpha \beta} U_{R, \alpha \beta}^*) \chi_-(y_\alpha, y_\beta)  \right] \nonumber \\
&- & \sum_{i<j} \left[ \left( |D_{L, ij}|^2  + |D_{R, ij}|^2 \right) \chi_+ (y_i, y_j ) + 2  {\mathcal Re}(D_{L, ij} D_{R, ij}^*) \chi_-(y_i, y_j)  \right]  \Big] \label{ES}
\eeq
where $(\alpha, \beta)$ specify the up-type quarks and $(i, j)$ are for the down-type quarks. This generalizes the result in Ref.\cite{Lavoura:1992np}.

\section{Model detail}\label{Appendix2}
We  first report the gauge-fermion couplings  that are relevant to the $S,T, U$ calculation. Note that the couplings below are not rotated into the mass basis. For the fiveplet $\psi_5$,  the gauge interaction is constructed from the CCWZ object, but we can  separate out the part without misalignment to be:
\beq
&& Tr \left[\bar \psi_5 \gamma^\mu \left(V_\mu \psi_5 + \psi_5 V_\mu^T \right) + g_1 B_\mu^0 \hat X \bar \psi_5 \gamma^\mu \psi_5  \right]     \nonumber \\ 
   &=&  \frac{g_2}{2} \Big[ W_1^\mu \left( \bar{T} \gamma_\mu B+\bar{X} \gamma_\mu T_X \right) -  i  W_2^\mu \left( \bar{T} \gamma_\mu B + \bar{X} \gamma_\mu T_X\right) + h.c.  \nonumber \\ &  + & W_3^\mu \left( \bar{T} \gamma_\mu T -\bar{B} \gamma_\mu B  + \bar{X} \gamma_\mu X - \bar{T}_{X} \gamma_\mu T_{X} \right)   \nonumber \\ &  + & \frac{1}{6}  B_0^\mu \tan (\text{$\theta $w}) \left( \bar{T} \gamma_\mu T + \bar{B} \gamma_\mu B  + 7 \bar{X} \gamma_\mu X + 7 \bar{T}_{X} \gamma_\mu T_{X} \right)
\Big]
\eeq
with $V_\mu = g_2 W_\mu^i T_L^i + g_1 B_\mu T_R^3$.  The pure misaligned term  encoded in the  $\delta E_\mu = E_\mu -V_\mu$ part can be expanded to be:
\beq
 &&  Tr \left[\bar  \psi_5 \gamma^\mu  \left( \delta E_\mu \psi_5 + \psi_5 \delta E_\mu^T \right) \right]   \supset \nonumber \\ 
 & - & \frac{g_2 \sin ^2\left(\frac{\alpha }{2}\right)}{2} \Big[ W_1^{\mu } \left(\bar{B}+\bar{X}\right) \gamma _{\mu } \left(T+ T_X\right) + i W_2^\mu \left(\bar{B}-\bar{X}\right) \gamma _{\mu } \left(T+ T_X\right)\nonumber \\
 & + &  2 W_3^\mu \left( \bar{T} \gamma_\mu T - \bar{T}_{X} \gamma_\mu T_{X} \right)  - 2 B_0^\mu \tan (\text{$\theta $w}) \left( \bar{T} \gamma_\mu T - \bar{T}_{X} \gamma_\mu T_{X} \right)  \Big]     + h.c. \, \label{Eterm}
\eeq
The $d_\mu$ term without the pNGB derivative coupling  is:
\beq
Tr[ \bar \psi_5  d_\mu \gamma^\mu \psi_1] + h.c.  & \supset & \frac{g_2 \sin (\alpha )}{2 \sqrt{2}}   \Big[  W_1^\mu  \left(\bar{B}+\bar{X}\right)  + i  W_2^\mu \left(\bar{B}-\bar{X}\right)  \nonumber \\ &+&  W_3^\mu \left(\bar{T}-\bar{T}_{X}\right) - B_0^\mu \tan (\text{$\theta $w}) \left(\bar{T}-\bar{T}_{X}\right)  \Big] \gamma_\mu T_1  + h.c. \, \label{dterm}
\eeq

Let us consider  the SM  $(t_L, b_L)$ and $t_R$ are embedded in the adjoint spurion,  the top partners in $\psi_5$ and elementary fermions can be arranged into up and down sectors:
\beq
\mathcal{U} \equiv \left( t, T, T_X, \tilde{T} \right)^T \, \quad  \mathcal{D} \equiv \left( b, B  \right)^T
\eeq
To diagonalize the up and down masses, i.e. $\Omega^\dag_L M_{2/3} \Omega_R = M_{2/3}^{diag}$ and  $\Omega^{d \dag}_L M_{-1/3} \Omega_R^d = M_{-1/3}^{diag}$, the rotation for one bi-doublet scenario in Eq.(\ref{AdjMass}) at $\mathcal{O}(\epsilon^2)$ ($\epsilon \equiv \sin \alpha$) is :
{\footnotesize
\beq
& &  \Omega_L  = \left(
\begin{array}{cccc}
 \frac{M_5}{\sqrt{M_5^2+f^2 y_{L1}^2}} & \frac{f y_{L1}}{\sqrt{M_5^2+f^2
   y_{L1}^2}} & 0 & 0 \\
- \frac{f y_{L1}}{\sqrt{M_5^2+f^2 y_{L1}^2}} & \frac{M_5}{\sqrt{M_5^2+f^2
   y_{L1}^2}} & 0 & 0 \\
 0 & 0 & 1 & 0 \\
 0 & 0 & 0 & 1 \\
\end{array} 
\right)   \\ \nonumber \\
&+ & \epsilon^2 \left(
\begin{array}{cccc}
 \frac{ M_5 f^2  y_{L1}^2
   \left(f^2 \left(y_{L1}^2+y_{R1}^2\right) +M_5^2\right)}{4 \left(M_5^2+f^2
   y_{L1}^2\right)^{5/2}} & -\frac{ M_5^2 f  y_{L1}
   \left( f^2 \left(y_{L1}^2+y_{R1}^2\right) +M_5^2\right)}{4 \left(M_5^2+f^2
   y_{L1}^2\right)^{5/2}} & \frac{f  y_{R1}^2}{4 M_5  y_{L1}} & 0 \\
 \frac{ M_5^2 f y_{L1} \left( f^2 \left(y_{L1}^2+y_{R1}^2\right) +M_5^2\right)}{4  \left(M_5^2+f^2
   y_{L1}^2\right)^{5/2}} & \frac{ M_5 f^2  y_{L1}^2 \left( f^2 \left(y_{L1}^2+y_{R1}^2\right)+M_5^2\right)}{4 \left(M_5^2+f^2
   y_{L1}^2\right)^{5/2}}  & \frac{1}{4} \left(\frac{y_{R1}^2}{y_{L1}^2} -\frac{f^2 y_{R1}^2}{M_5^2} -1\right) & 0 \\
 -\frac{\left(M_5^2+f^2 y_{R1}^2\right)}{4 M_5^2 \sqrt{\frac{M_5^2}{f^2
   y_{L1}^2}+1}} & \frac{  \left(y_{L1}^2-y_{R1}^2\right)}{4 y_{L1}^2 \sqrt{\frac{f^2
   y_{L1}^2}{M_5^2}+1}} & 0 & 0 \\
 0 & 0 & 0 & 0 \\
\end{array}
\right)  \nonumber
\eeq
\beq
\Omega_R = \left(
\begin{array}{cccc}
1- \frac{ \epsilon ^2 \left(\frac{M_5^4}{\left(M_5^2+f^2
   y_{L1}^2\right)^2}+1\right) f^2 y_{R1}^2}{8 M_5^2} & -\frac{ \epsilon 
   M_5 f y_{R1}}{2 \left(M_5^2+f^2 y_{L1}^2\right)} & \frac{ \epsilon 
   f y_{R1}}{2 M_5} & 0 \\
 \frac{\epsilon  M_5 f  y_{R1}}{2 \left(M_5^2+f^2 y_{L1}^2\right)} &
   1-\frac{\epsilon ^2 M_5^2  f^2  y_{R1}^2}{8 \left(M_5^2+f^2
   y_{L1}^2\right)^2} & \frac{\epsilon ^2}{4} 
   \left(\frac{y_{R1}^2}{y_{L1}^2}-1\right) & 0 \\
 -\frac{ \epsilon  f y_{R1}}{2 M_5} & \frac{ \epsilon ^2}{4} \left(1-\frac{M_5^2
   y_{R1}^2}{ y_{L1}^2 \left(M_5^2+f^2 y_{L1}^2\right)}\right) &
   1-\frac{ \epsilon ^2 f^2 y_{R1}^2}{8 M_5^2} & 0 \\
 0 & 0 & 0 & 1 
\end{array}
\right)
\eeq 
\beq
&& \Omega_L^d = \left(
\begin{array}{cc}
 \frac{M_5}{\sqrt{M_5^2+f^2 y_{L1}^2}} & \frac{f y_{L1}}{\sqrt{M_5^2+f^2
   y_{L1}^2}} \\
 -\frac{f y_{L1}}{\sqrt{M_5^2+f^2 y_{L1}^2}} & \frac{M_5}{\sqrt{M_5^2+f^2
   y_{L1}^2}} \\
\end{array}
\right)  +
\epsilon ^2 \left(
\begin{array}{cc}
 \frac{ M_5 f^2  y_{L1}^2}{2 \left(M_5^2+f^2 y_{L1}^2\right)^{3/2}} & -\frac{  M_5^2 f y_{L1}}{2
   \left(M_5^2+f^2 y_{L1}^2\right)^{3/2}} \\
 \frac{ M_5^2 f  y_{L1}}{2 \left(M_5^2+f^2 y_{L1}^2\right)^{3/2}} & \frac{  M_5 f^2 y_{L1}^2}{2 \left(M_5^2+f^2 y_{L1}^2\right)^{3/2}} 
\end{array}
\right)
\eeq }
and $\Omega^d_R =1_{2\times 2}$. The rotation  for one singlet scenario in Eq.(\ref{AdjMass}) is:
{\footnotesize
\beq
&& \Omega_L =\left(
\begin{array}{cccc}
 1-\frac{ \epsilon ^2 M_5^2 f^2 y_{L2}^2}{4 \left(M_5^2+f^2
   y_{R2}^2\right)^2} & 0 & 0 & - \frac{ \epsilon  M_5 f y_{L2}}{\sqrt{2}
   \left(M_5^2+f^2 y_{R2}^2\right)} \\
 0 & 1 & 0 & 0 \\
 0 & 0 & 1 & 0 \\
 \frac{\epsilon  M_5 f  y_{L2}}{\sqrt{2} \left(M_5^2+f^2 y_{R2}^2\right)} &
   0 & 0 & 1-\frac{ \epsilon ^2 M_5^2 f^2 y_{L2}^2}{4 \left(M_5^2+f^2
   y_{R2}^2\right)^2} 
\end{array}
\right) 
\eeq
\beq
& &   \Omega_R = \left(
\begin{array}{ccccc}
 \frac{M_5}{\sqrt{M_5^2+f^2 y_{R2}^2}} & 0 & 0 & \frac{f
   y_{R2}}{\sqrt{M_5^2+f^2 y_{R2}^2}}  \\
 0 & 1 & 0 & 0   \\
 0 & 0 & 1 & 0  \\
 -\frac{f y_{R2}}{\sqrt{M_5^2+f^2 y_{R2}^2}} & 0 & 0 &
   \frac{M_5}{\sqrt{M_5^2+f^2 y_{R2}^2}} 
\end{array}
\right)  \nonumber  \\ \nonumber \\
&+& \epsilon ^2 \left(
 \begin{array}{cccc}
 \frac{  M_5^3 f^2 y_{R2}^2
   \left( f^2 \left(y_{L2}^2+y_{R2}^2\right)+M_5^2\right)}{2 \left(M_5^2+f^2
   y_{R2}^2\right)^{7/2}} & 0 & 0 & -\frac{ M_5^4 f  y_{R2}
   \left(f^2 \left(y_{L2}^2+y_{R2}^2\right) +M_5^2\right)}{2 \left(M_5^2+f^2
   y_{R2}^2\right)^{7/2}} \\
 0 & 0 & 0 & 0 \\
 0 & 0 & 0 & 0 \\
 \frac{  M_5^4 f y_{R2} \left( f^2 \left(y_{L2}^2+y_{R2}^2\right)+M_5^2\right)}{2 \left(M_5^2+f^2
   y_{R2}^2\right)^{7/2}} & 0 & 0 & \frac{  M_5^3 f^2 y_{R2}^2 \left(f^2 \left(y_{L2}^2+y_{R2}^2\right) +M_5^2\right)}{2 \left(M_5^2+f^2
   y_{R2}^2\right)^{7/2}} 
\end{array}
\right) 
\eeq }
Note that the rotation conserves the unitarity: $\Omega_{L}^\dag \Omega_L = \Omega_{L}^{d \dag} \Omega_L^d  = \Omega_{R} \Omega_R^\dag = 1 + \mathcal{O}(\epsilon^3)$.

We listed the  generators for the $SU(4)/Sp(4)$ model below,  in particular $X^{1, \cdots 4}$ forms a bi-doublet in $SU(2)_L \times SU(2)_R$.
\beq
&& ~~  S^{i} = \frac{1}{2} \left(  \begin{array}{cc}
\sigma_i & 0 \\
0 & 0
\end{array} \right)\,, \qquad \qquad   S^{i+3} = \frac{1}{2} \left(  \begin{array}{cc}
0 & 0 \\
0 & - \sigma_i^T 
\end{array} \right)\,,  \quad i = \{ 1, 2, 3\} 
\\
& &  S^{7} = \frac{i}{2\sqrt{2}} \left(  \begin{array}{cc}
0 & \sigma_3  \\
- \sigma_3 & 0 \end{array} \right)\,, \qquad S^{8} = \frac{1}{2\sqrt{2}} \left(  \begin{array}{cc}
0 & 1_2 \\
1_2 & 0 
\end{array} \right)\,, 
\\
& & S^{9} = \frac{i}{2\sqrt{2}} \left(  \begin{array}{cc}
0 & \sigma_1  \\
- \sigma_1 & 0 \end{array} \right)\,, \qquad S^{10} = \frac{i}{2\sqrt{2}} \left(  \begin{array}{cc}
0 & \sigma_2 \\
-\sigma_2 & 0 
\end{array} \right)\,. 
\eeq

\beq
& X^1 = -  \frac{1}{2 \sqrt{2}} \left( \begin{array}{cc}
0 & \sigma_3 \\
\sigma_3 & 0
\end{array} \right)\,, \quad  X^2 = \frac{i}{2 \sqrt{2}} \left( \begin{array}{cc}
0 & 1_2 \\
- 1_2 & 0
\end{array} \right)\,, \quad  X^3 = \frac{1}{2 \sqrt{2}} \left( \begin{array}{cc}
0 & \sigma_1 \\
\sigma_1 & 0
\end{array} \right)\,,  & \\
&  X^4 = \frac{1}{2 \sqrt{2}} \left( \begin{array}{cc}
0 & \sigma_2 \\
\sigma_2 & 0
\end{array} \right)\,, \quad 
X^5 = \frac{1}{2 \sqrt{2}} \left( \begin{array}{cc}
1_2 & 0 \\
 0 & - 1_2
\end{array} \right)\,. &
\eeq
For the $SU(4)/Sp(4)$ model, the CCWZ objects can be exactly evaluated. In the original basis, we find the following identity:
\beq
U_{\hat \Pi} =  \exp \left( i \frac{  \sqrt{2}\hat \Pi}{f} \right)  = \cos \frac{\sqrt{h^2+ \eta^2}}{2f} + i \frac{2 \sqrt{2} f}{\sqrt{h^2+ \eta^2}} \sin \frac{\sqrt{h^2+ \eta^2}}{2f}  \frac{\hat \Pi}{f}
\eeq
The misalignment is generated by the rotation $ U_\Pi = U_\alpha U_{\hat \Pi} U^{-1}_\alpha$. Then projecting  $i \,U_\Pi^{-1} D_\mu U_\Pi $ into the unbroken and broken directions, we obtain:
\beq
E_\mu &=& \sum_i^3 \left( g_2 W_\mu^i  S^i + g_1 B_\mu S^6 \right) +  \sum_i^3 \left( g_2 W_\mu^i -g_1 B_\mu \delta^{i 3 }\right)  \nonumber \\& & \left[
 \left(\frac{\cos \alpha  \left(h^2 \cos \frac{\sqrt{h^2 + \eta^2}}{f}+\eta^2\right)}{h^2+\eta^2}- \frac{h \sin \alpha  \sin \frac{\sqrt{h^2+\eta ^2}}{f}}{\sqrt{h^2+\eta^2}} - 1\right) (S^i- S^{i+3}) \right. \nonumber \\ &+& \left. \frac{ \eta }{\sqrt{2}}  \left( \frac{\sin \alpha \sin \frac{\sqrt{h^2+\eta^2}}{f}}{ \sqrt{h^2+\eta^2} }+ \frac{2 h \cos \alpha  \sin ^2 \frac{\sqrt{h^2 + \eta^2}}{2f}}{ \left(h^2 + \eta^2\right)} \right)  S^{i+6} \right] \nonumber \\ 
   &+& \sqrt{2}  \left(\cos \frac{\sqrt{h^2 + \eta^2}}{f}-1\right) \frac{h \partial_\mu \eta - \eta \partial_\mu h }{h^2 + \eta^2} S^{10}
\eeq
\beq
d_\mu &=& \sum_i^3 \left( g_2 W_\mu^i -g_1 B_\mu \delta^{i 3 }\right) \left( \sin \alpha  \cos \frac{\sqrt{h^2 + \eta^2}}{f}+\frac{h \cos (\alpha ) \sin \frac{\sqrt{h^2 + \eta^2}}{f}}{\sqrt{h^2 + \eta^2}} \right) X^i \nonumber \\
&+& \sqrt{2} \left(- \frac{h}{ 2 f}  \frac{\partial_\mu(h^2+ \eta^2)}{h^2 + \eta^2 } +   \frac{\eta \left(h \partial_\mu \eta - \eta \partial_\mu h \right) }{(h^2 + \eta^2)^{3/2}} \sin \frac{\sqrt{h^2 + \eta^2}}{f} \right) X^4  \nonumber \\
&+& \sqrt{2} \left(- \frac{\eta}{ 2 f}  \frac{\partial_\mu(h^2+ \eta^2)}{h^2 + \eta^2 } +   \frac{h \left(\eta \partial_\mu h - h \partial_\mu \eta \right) }{(h^2 + \eta^2)^{3/2}} \sin \frac{\sqrt{h^2 + \eta^2}}{f} \right) X^5
\eeq

\section{Top and bottom spurions}  \label{Appendix3}

The SM top and bottom quarks are put in the incomplete $\mathcal{G} = SU(4)$ representations.  For the antisymmetric and symmetric embedding,  the spurions of  $(t_L, b_L)$  and $t_R$ are: 
 \begin{eqnarray}
\footnotesize
A_{L} = 
\begin{pmatrix}
0 & 0 &  \frac{t_L}{\sqrt{2}} & 0  \\
0 & 0 &   \frac{b_L}{\sqrt{2}} & 0  \\
-\frac{t_L}{\sqrt{2}} & - \frac{b_L}{\sqrt{2}} & 0 & 0  \\
0 & 0 & 0 & 0 
\end{pmatrix}
\qquad 
S_{L} =
\begin{pmatrix}
0 & 0 &  \frac{t_L}{\sqrt{2}} & 0  \\
0 & 0 &   \frac{b_L}{\sqrt{2}} & 0  \\
\frac{t_L}{\sqrt{2}} &  \frac{b_L}{\sqrt{2}} & 0 & 0  \\
0 & 0 & 0 & 0 
\end{pmatrix}
\end{eqnarray}	

\beq
A_{R}  = \frac{i}{2} \begin{pmatrix}
\sigma_{2}  & 0   \\
   0 &  - \sigma_{2}   \\
\end{pmatrix} t_R
\qquad 
A_{R}^{(2)} = \frac{i}{2} \begin{pmatrix}
\sigma_2  & 0   \\
   0 &   \sigma_2  \\
\end{pmatrix} t_R
\qquad 
S_{R} = \frac{1}{\sqrt{2}} \begin{pmatrix}
0  & 0   \\
   0 &   \sigma_1  \\
\end{pmatrix}  t_R
\eeq 
In $SU(4)$, the adjoint representation is  a $\bf{15}$-plet, decomposing into ${\bf{10}}_S \oplus {\bf{5}}_A$ in the  $\mathcal{H} = Sp(4)$  subgroup. And the corresponding spurions for top and bottoms are:
 \begin{eqnarray}
\footnotesize
D_{L, A} = A_{L} \Sigma_B = 
\begin{pmatrix}
0 & 0 & 0  & - \frac{t_L}{\sqrt{2}}   \\
0 & 0 &  0 &  - \frac{b_L}{\sqrt{2}}   \\
\frac{b_L}{\sqrt{2}} & - \frac{t_L}{\sqrt{2}} & 0 & 0  \\
0 & 0 & 0 & 0 
\end{pmatrix}
\qquad 
D_{L, S} = S_{L} \Sigma_B =
\begin{pmatrix}
0 & 0 & 0 & - \frac{t_L}{\sqrt{2}}   \\
0 & 0 & 0 & -  \frac{b_L}{\sqrt{2}}  \\
- \frac{b_L}{\sqrt{2}} &  \frac{t_L}{\sqrt{2}} & 0 & 0  \\
0 & 0 & 0 & 0 
\end{pmatrix}
\end{eqnarray}	

\beq
D_{R, S} = S_{R} \Sigma_B = \frac{1}{\sqrt{2}} \begin{pmatrix}
0  & 0   \\
   0 &   \sigma_3  \\
\end{pmatrix}  t_R
\qquad 
D_{R, A} = A_{R}^{(2)} \Sigma_B= - \frac{1}{2} \begin{pmatrix}
\mathbbm{1}_{2}  & 0   \\
   0 &  - \mathbbm{1}_{2}   \\
\end{pmatrix}  t_R
\eeq

\bibliographystyle{JHEP}

\bibliography{CHMtop.bib}

\end{document}